\documentclass[10pt,a4paper]{article}

\usepackage[top=1.in, bottom=1.in, left=0.9in, right=0.9in]{geometry}

\usepackage[utf8]{inputenc} % Support for more character glyphs
\usepackage[T1]{fontenc}
\usepackage{graphicx} % Required to include images
\usepackage{color} % Required for custom colors
\usepackage{amsmath,amssymb,theorem} % Math packages
\usepackage{listings} % Required for including snippets of code
\usepackage{booktabs} % Required for better horizontal rules in tables
\usepackage{rotating} % Allows tables and figures to be rotated
\usepackage{microtype} % Slightly tweak font spacing for aesthetics
\usepackage{caption}
\usepackage{sectsty}
\usepackage[round,authoryear]{natbib}
\usepackage{mathtools}
\usepackage{float}
\usepackage{enumerate}

\usepackage{bm}
\usepackage{verbatim}
\usepackage{graphicx}

\usepackage{authblk}
\usepackage{hyperref}
\allsectionsfont{\bfseries\sffamily}

\usepackage[dvipsnames]{xcolor}

\date{}

\newcommand{\J}{\textbf{J}}
\newcommand{\JS}{\textbf{J}_S}
\newcommand{\deriv }{\mathrm{d}}
\newcommand{\ic}{\textbf{r}_0}
\newcommand{\I}{\textbf{I}}
\newcommand{\lmax}{\lambda_{\rm max}}
\newcommand{\prop}{\textbf{P}_t}
\newcommand{\propp}{\textbf{P}_{\tstar}}

\newcommand{\Tr}{{\rm Tr}}
\newcommand{\deter}{{\rm Det}}
\newcommand{\LJp}{\lambda^+}
\newcommand{\LJm}{\lambda^-}
\newcommand{\LJpm}{\lambda^\pm}
\newcommand{\LJSp}{\lambda^+_S}
\newcommand{\LJSm}{\lambda^-_S}
\newcommand{\LJSpm}{\lambda^\pm_S}
\newcommand{\tstar}{t^*}
\newcommand{\uv}{\textbf{u}\textbf{v}^T}
\newcommand{\dd}{{\rm d}}
\newcommand{\diag}{\textbf{diag}}
\newcommand{\argmax}{{\rm argmax}}

\title{\sf \LARGE Coding with transient trajectories in recurrent neural networks}
\author[1]{Giulio Bondanelli}
\author[2]{Srdjan Ostojic}
\affil[1]{\small something}
\affil[2]{something}

\definecolor{niceblue}{RGB}{65,105,225}
\definecolor{deepcarmine}{rgb}{0.66, 0.13, 0.24}
\definecolor{nicegrey}{RGB}{30,83,132}

\hypersetup{
	colorlinks=true,       % false: boxed links; true: colored links
	linkcolor=black,          % color of internal links (change box color with linkbordercolor)
	citecolor=nicegrey,        % color of links to bibliography
}

\begin{document}
\begin{center}
\LARGE{\sf Coding with transient trajectories in recurrent neural networks}

\vspace{0.5cm}
\large{Giulio Bondanelli \textsuperscript{1}, Srdjan Ostojic \textsuperscript{1}}
\normalsize{
	\\
	\bigskip
	\textsuperscript{1} Laboratoire de Neurosciences Cognitives et Computationelles, D\'epartement d’\'etudes cognitives, ENS, PSL University, INSERM, Paris, France
	\\
	
        % \textsuperscript{\bf*} Lead author
        }

\end{center}

\vspace*{0.5cm}
\section*{Abstract }

Following a stimulus, the neural response typically strongly varies in
time  and across  neurons before  settling to  a  steady-state.  While
classical population coding  theory disregards the temporal dimension,
recent works  have argued that trajectories of  transient activity can
be particularly  informative about stimulus identity and  may form the
basis of computations through  dynamics.  Yet the dynamical mechanisms
needed to  generate a population code based  on transient trajectories
have not been fully elucidated.  Here we examine transient coding in a
broad  class  of   high-dimensional  linear  networks  of  recurrently
connected units.  We start by reviewing a well-known result that leads
to a distinction  between two classes of networks:  networks in which
all inputs  lead to weak,  decaying transients, and networks  in which
specific inputs elicit strongly  amplified transient responses and are
mapped onto  orthogonal output states during the  dynamics. Theses two
classes  are  simply  distinguished  based  on  the  spectrum  of  the
symmetric part  of the connectivity  matrix.  For the second  class of
networks, which  is a sub-class  of non-normal networks, we  provide a
procedure   to   identify  transiently   amplified   inputs  and   the
corresponding  readouts.  We  first  apply these  results to  standard
randomly-connected  and   two-population  networks.   We   then  build
minimal,  low-rank  networks   that  robustly  implement  trajectories
mapping a  specific input  onto a specific  output state.  Finally, we
demonstrate  that  the capacity  of  the  obtained networks  increases
proportionally with their size.

\vspace*{0.5cm}
\section*{Significance statement }

 Classical theories of sensory coding consider the neural activity following a stimulus as constant in time. Recent works have however suggested that the temporal variations following the appearance and disappearance of a stimulus are strongly informative. Yet their dynamical origin remains little understood. Here we show that strong temporal variations in response to a stimulus can be generated by collective interactions within a network of neurons if the connectivity between neurons satisfies a simple mathematical criterion. We moreover determine the relationship between connectivity and the stimuli that are represented in the most informative manner by the variations of activity, and estimate the number of different stimuli a given network can encode using temporal variations of neural activity.

\newpage

\section*{Introduction}\
The brain represents sensory stimuli in terms of the collective activity of thousands of neurons. Classical population coding theory describes the relation between stimuli and neural firing in terms of tuning curves, which assign a single number to each neuron in response to a stimulus \citep{Seung1993, pouget_dayan_zemel_2000, pouget_dayan_zemel_2003}. The activity of a neuron following a stimulus presentation typically strongly varies in time and  explores a range of values, but classical population coding typically leaves out such dynamics by considering either time-averaged or steady-state firing.

In contrast to this static picture, a number of recent works have argued that the temporal dynamics of population activity may play a key role in neural coding and computations \citep{Rabinovich2008, Rabinovich2008_PlosCB,  Durstewitz2008, Buonomano2009, Brody2003, Crowe2010, Jun2010, Shafi2007, Laje2013, Chaisangmongkon2017, Goudar2017}. As the temporal response to a stimulus is different for each neuron, an influential approach has been to represent population dynamics in terms of temporal trajectories in the neural state space, where each axis corresponds to the activity of one neuron \citep{Churchland2007,Mazor2005, Machens2010,Mante2013}. Coding in this high-dimensional space is typically examined by combining linear decoding and dimensionality-reduction techniques \citep{Cunningham2014, Machens2016, Bagur2018}, and the underlying network is often conceptualised in terms of a dynamical system \citep{Shenoy2013, Churchland2010, Churchland2012, Michaels2016, Mante2013, Wang2018, Remington2018,
Hennequin2014, Carnevale2015, Sussillo2014}. Such approaches have revealed that the discrimination between stimuli based on neural activity can be higher during the transient phases than at steady state \citep{Mazor2005}, arguing for a coding scheme in terms of neural trajectories. A full theory of coding with transient trajectories is however currently lacking.

To produce useful transient coding, the trajectories of neural activity need to satisfy at least three requirements \citep{Rabinovich2008}. They need to be (i) stimulus-specific, (ii) robust to noise and (iii) non-monotonic, in the sense that the responses to different stimuli differ more during the transient dynamics than at steady-state. This third condition is crucial as otherwise coding with transients can be reduced to classical, steady-state population coding. Recent works have shown that recurrent networks with so-called non-normal connectivity can lead to amplified transients \citep{Ganguli2008, Murphy2009, Goldman2009, Hennequin2012, Hennequin2014, Ahmadian2015}, but sufficient conditions for such amplification were not given. We start by reviewing a well-known result linking the norm of the transient activity to the spectrum of the symmetric part of the connectivity matrix. This results leads to a simple distinction between two classes of networks: networks  in  which all  inputs  lead to  weak,
decaying  transients, and  networks  in which  specific inputs  elicit
strongly amplified transient responses.  We then characterize inputs that lead to non-monotonic trajectories,  and show that they induce transient dynamics that map inputs onto orthogonal output directions. We first apply these analyses to standard two-population and randomly-connected networks. We then specifically exploit these results to build low-rank connectivity matrices that implement specific trajectories to transiently encode specified stimuli, and examine the noise-robustness and capacity of this setup.

\section*{Results}\

We study  linear networks of $N$ randomly and recurrently coupled rate units with dynamics given by:

\begin{equation}\label{linear_rate_model}
\dot{r}_i=-r_i+\sum_{j=1}^NJ_{ij}r_j+I(t)r_{0,i}.
\end{equation}
Such networks can be interpreted as describing the linearized dynamics of a system around an equilibrium state. In this picture, the quantity $r_i$ represents the deviation of the activity of the unit $i$ from its equilibrium value, and  $J_{ij}$ denotes the effective strength of the connection from neuron $j$ to neuron $i$. Unless otherwise specified, we consider an arbitrary connectivity matrix $\J$.
Along with the recurrent input, each unit $i$ receives an external drive $I(t)r_{0,i}$ in which the temporal component $I(t)$ is equal for all neurons, and the vector $\ic$ (normalized to unity) represents the relative amount of input to each neuron.

\newpage
\subsection*{Monotonic vs. Amplified Transient trajectories}\

\begin{figure}[ht]
\centering
\includegraphics[scale=0.95]{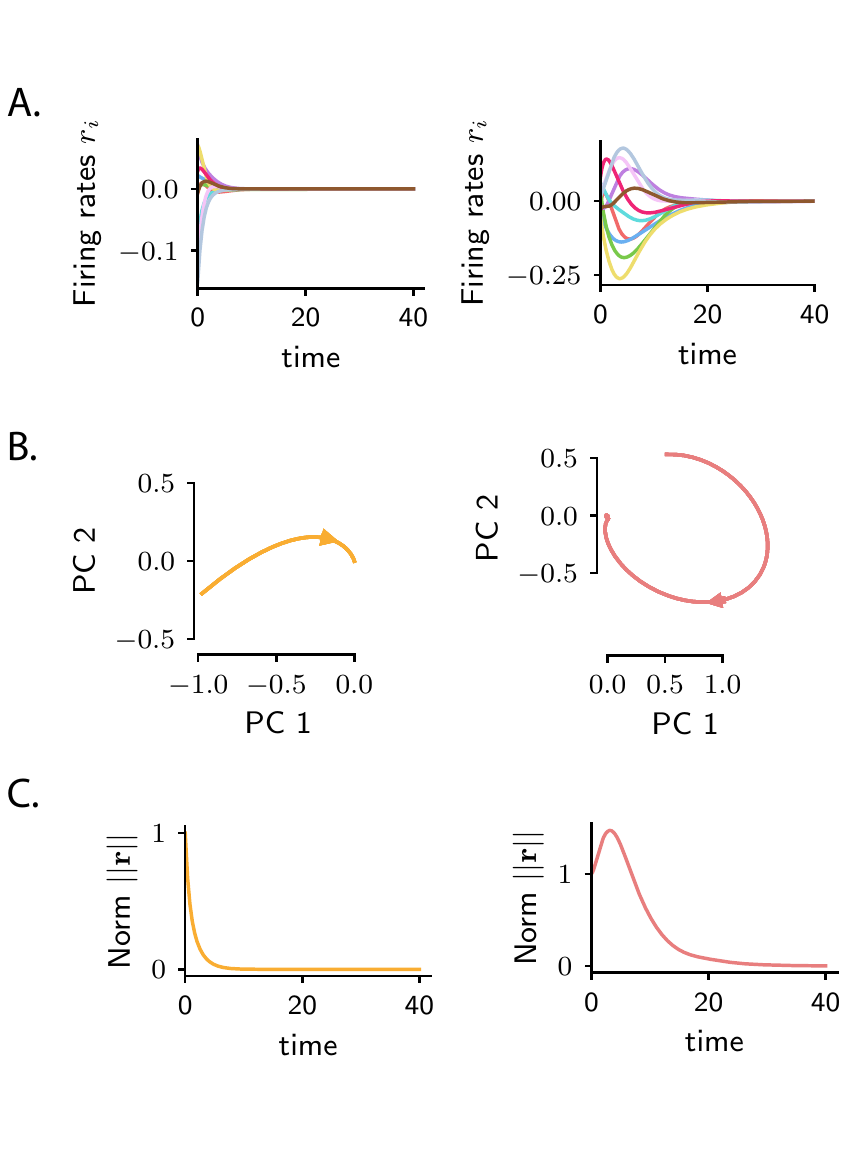} 
\caption{\label{fig1}
\textsf{\bfseries Monotonically decaying vs. amplified transient dynamics.}
 Dynamics of a linear recurrent network in response to a short external perturbation along a given input direction $\ic$. The left and right examples correspond to two different connectivity matrices, where the connection strengths are independently drawn from a Gaussian distribution with zero mean and variance equal to $g^2/N$ (left: $g=0.5$; right:  $g=0.9$). \textsf{\bfseries A.} Firing rate dynamics of 10 individual units.   \textsf{\bfseries B.} Projections of the population activity onto the first two principal components of the dynamics. Yellow and red color correspond respectively to $g=0.5$ and $g=0.9$. 
\textsf{\bfseries C.} Temporal dynamics of the activity norm $||\textbf{r}(t)||$. \textit{Left}: in the case of weakly non-normal connectivity the activity norm displays monotonic decaying behaviour for any external input perturbation. \textit{Right}: for strongly non-normal connectivity,  specific stimuli generate a transient increase of the activity norm.  $N=200$ in simulations.
}
\end{figure}

We focus on the transient dynamics in the network following a brief input in time ($I(t)=\delta(t)$) along the  external input direction $\ic$, which is equivalent to setting the initial condition  to $\ic$. The temporal activity of the network in response to this input can be represented as a trajectory $\textbf{r}(t)$ in the high-dimensional space in which the $i$-th component is the firing rate of neuron $i$ at time $t$. We assume the network is stable, so that the trajectory asymptotically decays to the equilibrium state that corresponds to $r_i=0$. At intermediate times, depending on the connectivity matrix $\J$ and on the initial condition $\ic$, the trajectory can however exhibit two qualitatively different types of behavior: it can either monotonically decay towards the asymptotic state, exploring essentially a single dimension, or transiently move away from it by following a rotation (Fig.~\ref{fig1} A-B). We call these two types of trajectories respectively monotonic and amplified. 

The two types of transient trajectories can be distinguished by looking at the Euclidean  distance between the activity  at time point $t$ and the  asymptotic equilibrium state, given by the activity norm $||\textbf{r}(t)||=\sqrt{r_1(t)^2+r_2(t)^2+...+r_N(t)^2}$. Focusing on the  norm allows us to deal with a single scalar quantity instead of $N$ firing rates. Monotonic and amplified  transient trajectories respectively correspond to monotonically decaying and transiently increasing  $||\textbf{r}(t)||$ (Fig.~\ref{fig1} C). Note that a transiently increasing  $||\textbf{r}(t)||$ necessarily implies that the firing rate of at least one neuron shows a transient increase before decaying to baseline.

One approach to understanding how the connectivity matrix $\J$ determines the transient trajectory is to project the dynamics  on the basis formed by the right-eigenvectors $\{\textbf{v}_k\}$ of $\J$ \citep{DayanAbbott}. The component $\tilde{r}_k(t)$ along the $k-$th eigenmode decays exponentially and  the activity norm can be expressed as:

\begin{equation}\label{norm_and_correlations_between_eigenvectors}
||\textbf{r}(t)||=\sqrt{\sum_{k=1}^N\tilde{r}_k(t)^2 + 2\sum_{k>j} \tilde{r}_k(t)\tilde{r}_j(t) (\textbf{v}_k\cdot\textbf{v}_j)  }.
\end{equation}
If all  the eigenvectors $\textbf{v}_k$ are  mutually orthogonal, then
the squared activity norm is  a sum of squares of decaying exponentials,
and therefore a monotonically decaying function. Connectivity matrices
$\J$ with all orthogonal eigenvectors  are called normal
matrices, and  they  thus  generate only   monotonic transients. In particular, any symmetric matrix is normal.
On the  other hand, connectivity matrices for  which some eigenvectors are not mutually orthogonal are called \textit{non-normal} \citep{trefethenBook}. For such
matrices,     the    second    term     under the   square     root    in
Eq.~\eqref{norm_and_correlations_between_eigenvectors}   can   have
positive or negative sign, so that the norm cannot in general be written as the sum of
decaying   exponentials.   It is well known that  non-normal   matrices  can   lead  to
non-monotonic    transient    trajectories  \citep{Trefethen578, Ganguli2008, Murphy2009, Goldman2009, Hennequin2012, Hennequin2014, Ahmadian2015} . 

Nonetheless,  a  non-normal  connectivity   matrix  $\J$  is  just  a
 necessary,  but  not a  sufficient  condition  for  the existence  of
 transiently  amplified trajectories.  As will  be  illustrated below,
 having non-orthogonal  eigenvectors does not  guarantee the existence
 of  transiently  amplified  inputs.   This  raises  the  question  of
 identifying the sufficient conditions on the connectivity matrix $\J$
 and input $\ic$ for the transient trajectory to be amplified.  In the
 following, we  point out a simple criterion on  the connectivity matrix
 $\J$ for the existence of amplified trajectories, and show that it is
 possible to  identify the different  inputs giving rise  to amplified
 trajectories and estimate their number.

\subsection*{Two classes of non-normal connectivity}\

To distinguish between monotonic and amplified trajectories, we focus on the rate of change $\deriv ||\textbf{r}({t})||/\deriv t$ of the activity norm. For a monotonic trajectory, this rate of change is negative at all times, while for amplified trajectories it transiently takes positive values before becoming negative as the activity decays to the equilibrium value. Using this criterion, we can determine the conditions under which a network generates an amplified trajectory for at  at least one input $\ic$. Indeed, the rate of change of the activity norm satisfies (see \cite{Trefethen578,Caswell97})

\begin{equation}\label{eq_norm}
\frac{1}{||\textbf{r}||}\frac{\deriv||\textbf{r}||}{\deriv t}=\frac{\textbf{r}^T(\textbf{J}_S-\textbf{I})\textbf{r}}{||\textbf{r}||^2},\qquad 
\JS=\frac{\J+\J^T}{2}
\end{equation}
Here the matrix $\JS$ denotes the symmetric part of the connectivity matrix $\J$. The right hand side of Eq.~\eqref{eq_norm} is a Rayleigh quotient \citep{Horn_matrixAnalysis}. It reaches its maximum value when $\textbf{r}(t)$ is aligned with the eigenvector of $\JS$ associated with its largest eigenvalue, $\lmax(\JS)$, and the corresponding maximal rate of change of the activity norm is therefore  $\lmax(\JS)-1$.

Eq.~\eqref{eq_norm} directly implies that a necessary and sufficient condition for the existence of transiently amplified trajectories is  that the largest eigenvalue of the symmetric part $\JS$  be larger than unity, $\lmax(\JS)>1$ \citep{Trefethen578}.
If that is the case, choosing the initial condition along the eigenvector associated with $\lmax(\JS)$  leads to a positive rate of change of the activity norm at time $t=0$, and therefore generates a transient increase of the norm corresponding to
 an amplified trajectory, which shows the sufficiency of the criterion. Conversely, if a given input produces an amplified trajectory, at least one eigenvalue of $\JS$ is necessarily larger than one. If that were not the case, the right hand side of the equation for the norm would take negative values for all vectors $\textbf{r}(t)$, implying a monotonic decay of the norm. This demonstrates the necessity of the criterion.

The criterion based  on the symmetric part of  the connectivity matrix
allows  us to  distinguish two  classes of  connectivity  matrices: if
$\lmax(\JS)<1$  all  external   inputs  $\ic$  lead  to  monotonically
decaying  trajectories (non-amplifying connectivity); if  $\lmax(\JS)>1$  specific input  directions
lead to a  non-monotonic amplified activity norm (amplifying connectivity).  The  key point here
is that for a non-normal  connectivity matrix $\J$, the symmetric part
$\JS$  is  in general  different  from  $\J$.  The condition  for  the
stability of the system ($\mathfrak{Re}\lmax(\J)<1$) and the condition
for  transient   amplification  ($\lmax(\JS)>1$)  are   therefore  not
mutually exclusive, except in  the case of one-dimensional dynamics or
symmetric connectivity matrices.

\begin{figure}[ht]
\centering
\includegraphics[scale=0.95]{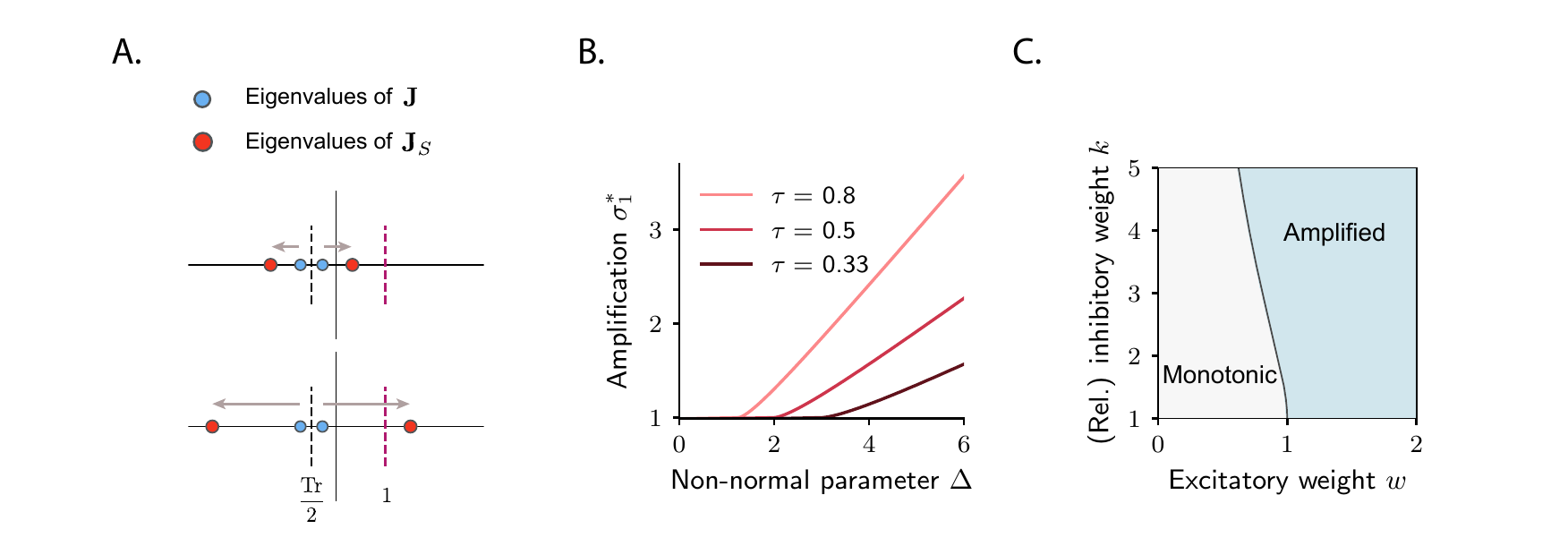} 
\caption{\label{fig:2pop}
\textsf{\bfseries Dynamical regimes for a network of two interacting populations.} \textsf{\bfseries A.} Relation between the eigenvalues of the connectivity matrix $\J$ (blue dots) and the eigenvalues of its symmetric part, $\JS$ (red dots). Both pairs of eigenvalues are symmetrically centered around the trace of $\J$, $\Tr(\J)$, but the eigenvalues of $\JS$ lie further apart (Eq.~\ref{eq:lambda-J-Js}), and the maximal eigenvalue of $\JS$ can cross unity if the difference $2\Delta$ between the off-diagonal elements of the connectivity matrix is sufficiently large (bottom panel).  \textsf{\bfseries B.} Value of the maximum amplification of the system (quantified by the maximal singular value $\sigma_1(\propp)$ of the propagator, see \textit{Methods}) as a function of the non-normal parameter $\Delta$. Here we fix the two eigenvalues of $\J$, the largest of which effectively determines the largest timescale of the dynamics, and vary $\Delta$. Colored traces correspond to different values of the largest timescale of the system $\tau=1/(1-\mathfrak{Re}\lmax(\J))$. For small values of $\Delta$ the maximum amplification is equal to one, and it increases approximately linearly when $\Delta$ is larger than the critical value. \textsf{\bfseries C.} Dynamical regimes for an excitatory-inhibitory two population model, as in \citep{Murphy2009}. Here $w$ represents the weights of the excitatory connections ($J_{EE}=J_{IE}=w$) and $-kw$ the weights of the inhibitory ones ($J_{EI}=J_{II}=-kw$, with the relative strength of inhibition $k>1$). In order to achieve  transient  amplification the excitatory weight $w$ has to be (approximately) larger than unity. In \textsf{\bfseries B.} each colored trace corresponds to a different choice of $\Tr(\J)$ and $\deter(\J)$. From top to bottom traces: $\Tr(\J)=-0.5,-2,-4$ and $\deter(\J)=\Tr^2(\J)/4$ (for convenience), corresponding respectively to $\tau=0.8, 0.5, 0.33$.
}
\end{figure}

The simplest illustration of this result is a two-population network. In that case the relationship between the eigenvalues of $\J$ and $\JS$ is straightforward. The eigenvalues of $\J$ and $\JS$  are given by
\begin{equation}\label{eq:lambda-J-Js}
\lambda^\pm (\J)=\frac{\Tr(\J)\pm\sqrt{\Tr^2(\J)-4\deter(\J)}}{2},\qquad  \lambda^\pm (\JS)=\frac{\Tr(\J)\pm\sqrt{\Tr^2(\J)-4\deter(\J)+4\Delta^2}}{2},
\end{equation}
where  $\Tr(\J)$ and $\deter(\J)$ are the trace and determinant of the full connectivity matrix $\J$,  and $2\Delta$ is the difference between the off-diagonal elements of $\J$. Assuming for simplicity that the eigenvalues of $\J$ are real, Eqs.~\eqref{eq:lambda-J-Js} show that the maximal eigenvalue of  $\JS$ is in general larger than the maximal eigenvalue of $\J$, and the difference between the two is controlled by the parameter $\Delta$ which quantifies how non-symmetric the matrix $\J$ is. If $\Delta$ is large enough, $\JS$ will have an unstable eigenvalue, even if both eigenvalues of  $\J$ are stable (Fig.~\ref{fig:2pop} A). The value of $\Delta$ therefore allows to distinguish between non-amplifying and amplifying connectivity. Furthermore, for amplifying connectivity, the  parameter $\Delta$ directly controls the amount of amplification in the network (Fig.~\ref{fig:2pop} B), defined as the maximum value of the norm $||\textbf{r}(t)||$ over time and initial conditions $\ic$ (see \textit{Methods}). A specific example is a network consisting of two interacting excitatory-inhibitory populations \citep{Murphy2009}. In that case our criterion states that the excitatory feedback needs to be (approximately) larger than unity in order to achieve transient amplification (Fig.~\ref{fig:2pop} C and \textit{Methods}).

\begin{figure}[ht]
\centering
\includegraphics[scale=0.95]{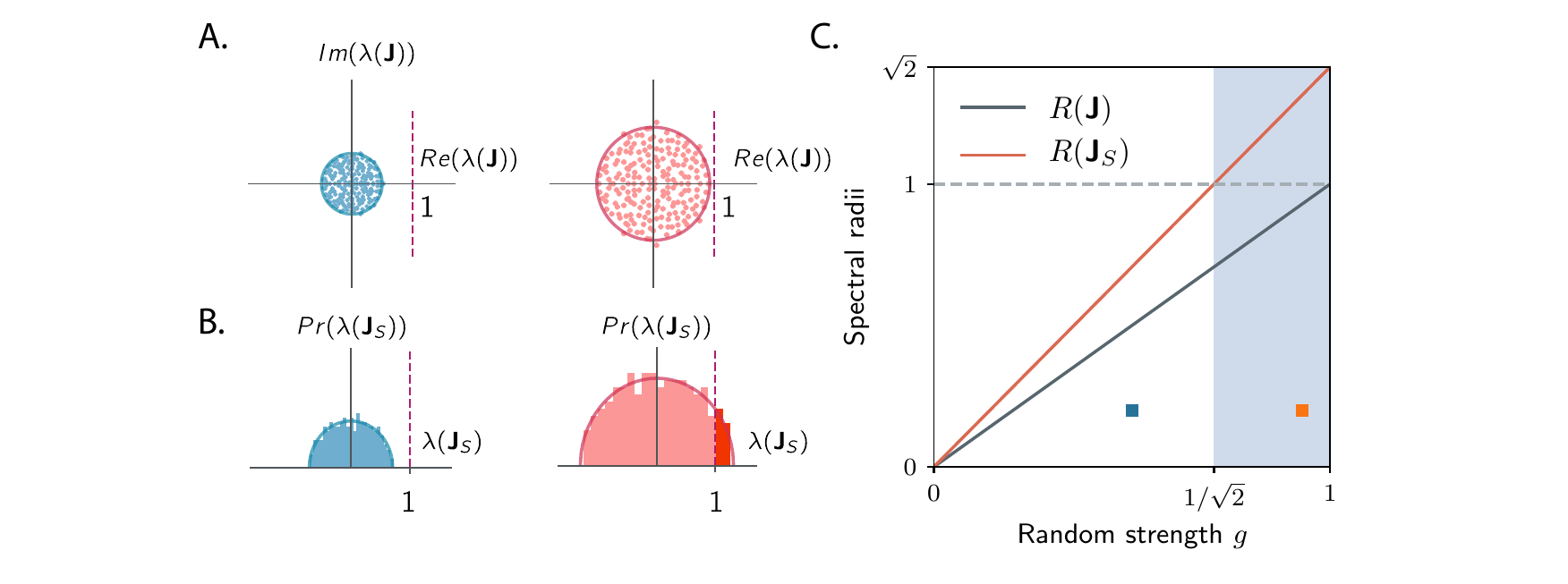} 
\caption{\label{fig:rand}
 \textsf{\bfseries Dynamical regimes of a $N$-dimensional network model with random Gaussian connectivity structure.}  Each entry of $\J$ is independently drawn from a Gaussian distribution with zero mean and variance  $g^2/N$. \textsf{\bfseries A.} The eigenvalues of $\J$ are complex, and, in the limit of large $N$,  distributed uniformly within a circle of radius $R(\J)=g$ in the complex plane (Girko's law, \citep{Girko}). The system is stable if $g<1$. Left: $g=0.5$. Right: $g=0.9$. 
\textsf{\bfseries B.} The eigenvalues of the symmetric part $\JS$ are real-valued, and   are distributed  in the large $N$ limit according to the semicircle law, with the largest eigenvalue of $\JS$ given by the spectral radius $R(\JS)=\sqrt{2}g$ \citep{Wigner55, Wigner58}. Since the spectral radius of $\JS$ is larger than the spectral radius of $\J$, for sufficiently large values of $g$  some eigenvalues of $\JS$ can be larger than unity (in red), while the network dynamics are stable ($g<1$).
\textsf{\bfseries C.} Spectral radii of $\J$ and $\JS$ as a function of the random strength $g$. The interval of values of $g$ for which the system displays strong transient dynamics in response to specific inputs is given by $1/\sqrt{2}<g<1$.  $N=200$ in simulations.
}
\end{figure}

A second illustrative example is a network of $N$ randomly connected neurons, where each connection strength is independently drawn from a Gaussian distribution with zero mean and variance equal to $g^2/N$. For such a network, the eigenvalues of $\J$ and $\JS$ are random, but their distributions are known.
The eigenvalues of $\J$ are uniformly distributed in the complex plane  on a circle of radius $g$ \citep{Girko}, so that the system is stable for $g<1$ (Fig.~\ref{fig:rand} B). On the other hand, the eigenvalues of the symmetric part $\JS$ are real and distributed according to the semicircle law with spectral radius $\sqrt{2}g$ \citep{Wigner55, Wigner58} (Fig.~\ref{fig:rand} B). The fact that the spectral radius of $\JS$ is larger by a factor $\sqrt{2}$ than the spectral radius of $\J$ implies that if $g$ is in the interval $1/\sqrt{2}<g<1$ the network is stable but exhibits amplified transient activity (Fig.~\ref{fig:rand} A). Note that the connectivity is non-normal for any value of $g$, but the additional condition $g>1/\sqrt{2}$ is needed for the existence of amplified trajectories. This in particular implies that for random connectivity transient amplification requires the network to be close to instability, so that the dynamics are slowed down as pointed out in \citep{Hennequin2012}.

\subsection*{Coding with amplified transients}

\begin{figure}[H]
\centering
\includegraphics[scale=0.95]{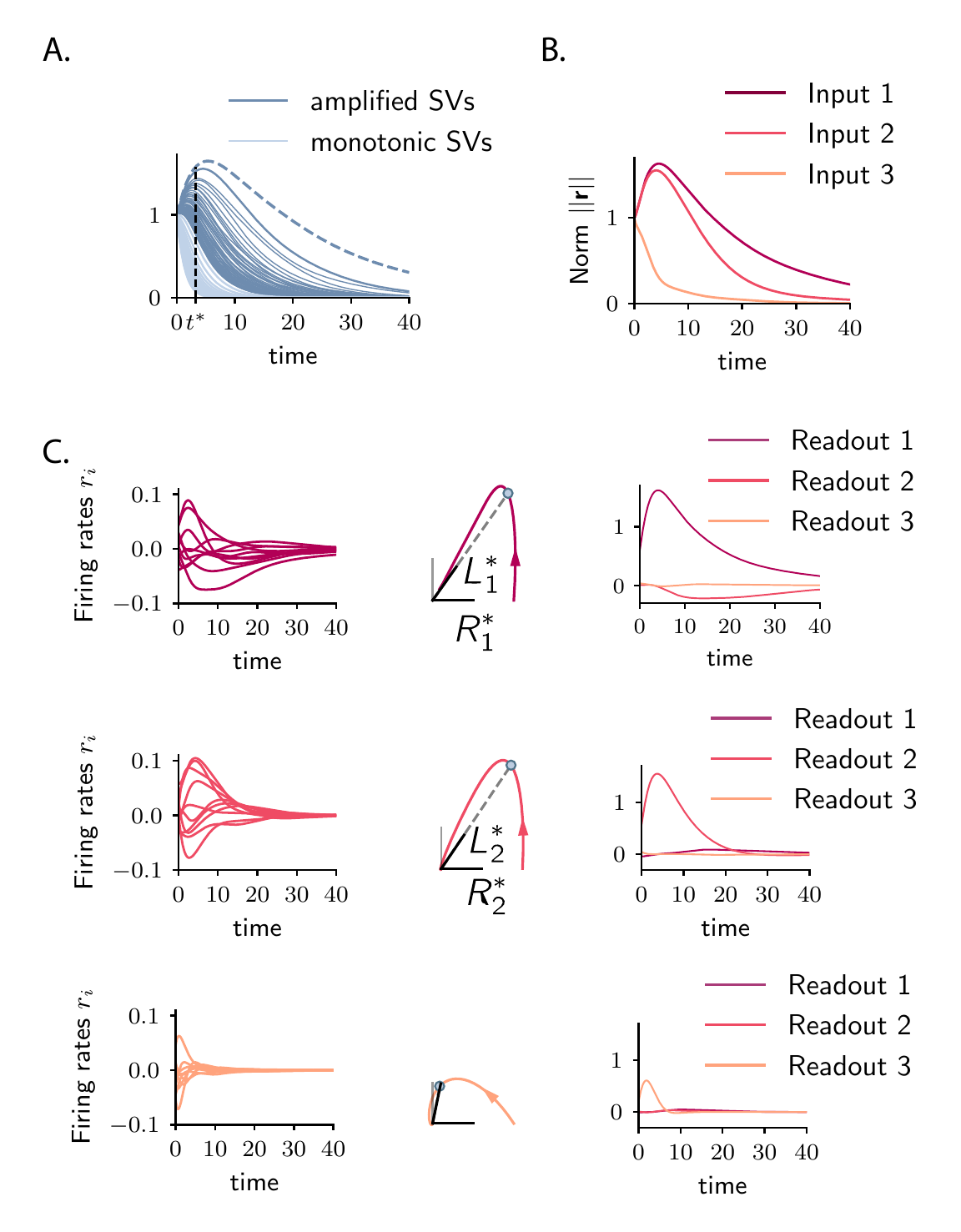} 
\caption{\label{fig:capacity}
\textsf{\bfseries Coding multiple stimuli with amplified transient trajectories.} Example corresponding to a $N$-dimensional Gaussian connectivity matrix with $g=0.9$.  \textsf{\bfseries A.} Singular values of the propagator, $\sigma_i^{(t)}$, as a function of time (SV trajectories). Dark blue traces show the amplified singular values, defined as having positive slope at time $t=0$; The dominant singular value $\sigma_1^{(t)}$ corresponds to the dashed line. Light blue traces correspond to the non-amplified singular values, having negative slope at $t=0$. \textsf{\bfseries B.} Norm of the activity elicited by the first two amplified inputs, i.e. $\textbf{R}_1^*$, $\textbf{R}_2^*$, (right singular vectors corresponding to singular values $\sigma_1^{(t^*)}$ and $\sigma_2^{(t^*)}$ at time $t^*$ in pannel A; purple and red traces), and by one non-amplified input (chosen as $\textbf{R}_{100}^*$, corresponding to $\sigma_{100}^{(t)}$; orange trace). \textsf{\bfseries C.} Illustration of the dynamics elicited by the three inputs as in {\bf B}. \textit{Left}: Activity of 10 individual units. \textit{Center}: Projections of the evoked trajectories onto the plane defined by the stimulus $\textbf{R}_i^*$ and the corresponding readout vector $\textbf{L}_i^*$ (in analogy with the amplified case, we chose the readout of the non-amplified dynamics to be the state of the system at time $\tstar$, i.e. $\textbf{L}_{100}^*$). \textit{Right}: population responses to the three stimuli projected on the readout vectors $\textbf{L}_1^*$, $\textbf{L}_2^*$ and $\textbf{L}_{100}^*$. 
$N=1000$ in simulations.
}
\end{figure}

For  a  connectivity  matrix  satisfying the  amplification  condition
$\lmax(\JS)>1$, only  specific external inputs $\ic$  are amplified by
the recurrent  circuitry, while others lead  to monotonically decaying trajectories
(Fig.~\ref{fig:capacity} B).  Which and how many  inputs  are
amplified? What  is the resulting state  of the network  at the time  of maximal
amplification, and how can the inputs be decoded from that state?

One approach to these questions  is to examine the mapping from inputs
to  states at a  given time  $t$ during the dynamics.
Since we consider linear networks,  the state reached at time $t$ from
the  initial   condition  $\ic$  is   given  by  the   linear  mapping
$\textbf{r}(t)=\prop\ic$,     where     for     any    time     $t>0$,
$\prop=\exp(t(\J-\I))$ is an $N \times N$ matrix called the propagator
of the network. At a  given time $t$, the singular value decomposition (SVD)
of $\prop$ defines a  set of singular values $\{\sigma_k^{(t)}\}$, and
two  sets   of  orthonormal  vectors   $\{\textbf{R}_k^{(t)}  \}$  and
$\{\textbf{L}_k^{(t)}\}$, such  that $\prop$ maps $\textbf{R}_k^{(t)}$
onto  $\sigma_k^{(t)}\textbf{L}_k^{(t)}$.    In  other  words,  taking
$\textbf{R}_k^{(t)}$ as the initial condition leads the network to the
state $\sigma_k^{(t)}\textbf{L}_k^{(t)}$ at time $t$:
\begin{equation}\label{SVD_propagator}
\textbf{r}(t)=\prop \textbf{R}_k^{(t)} =\sigma_k^{(t)} \textbf{L}_k^{(t)}.
\end{equation}
If $\sigma_k^{(t)}>1$, the norm of  the activity at time $t$ is larger
than  unity, so  that  the initial  condition $\textbf{R}_k^{(t)}$  is
amplified.  In fact, the  largest singular value of $\prop$ determines
the maximal  possible amplification at  time $t$ (see  \textit{Methods}).  Note
that  for  a  normal  matrix,  the left  and  right  singular  vectors
$\textbf{R}_k^{(t)}$ and  $\textbf{L}_k^{(t)}$ are identical,  and the
singular values are equal to the eigenvalues, so that the stability of
the  dynamics  imply  an absence  of  amplification.  Conversely,  stable
amplification      implies      that     $\textbf{R}_k^{(t)}$      and
$\textbf{L}_k^{(t)}$   are  not  identical,   so  that   an  amplified
trajectory explores at least two dimensions corresponding to the plane
spanned by $\textbf{R}_k^{(t)}$ and $\textbf{L}_k^{(t)}$.

Since  the propagator $\prop$  depends on  time, the  singular vectors
$\textbf{R}_k^{(t)}$ and  $\textbf{L}_k^{(t)}$,  and  the singular  values
$\sigma_k^{(t)}$  depend  on time.   One  can  therefore  look at  the
temporal trajectories $\sigma_k^{(t)}$,  which by definition all start
at  one at  $t=0$  (Fig.~\ref{fig:capacity} A).   If the  connectivity
satisfies  the condition  for  transient amplification,  at least  one
singular  value increases above  unity, and  reaches a  maximum before
asymptotically decreasing to zero.  The number of singular values that
simultaneously  take values  above  unity (Fig.~\ref{fig:capacity}  A)
defines the  number of orthogonal initial conditions  amplified by the
dynamics.   Choosing  a  time  $\tstar$  at  which  $N_s$  of  the  singular value
trajectories lie  above unity, we can  indeed identify a  set of $N_s$
orthogonal,  amplified  inputs  corresponding  to the  right  singular
vectors $\textbf{R}_k^{(\tstar)}$ of  the propagator at time $\tstar$.
According to Eq.~\eqref{SVD_propagator}, each of these inputs is mapped
in  an amplified  fashion to  the corresponding  left  singular vector
$\textbf{L}_k^{(\tstar)}$  at  time   $\tstar$,  which  also  form  an
orthogonal  set.  Each  amplified input  can therefore  be  decoded by
projecting  the network  activity on  the corresponding  left singular
vector  $\textbf{L}_k^{(\tstar)}$  (Fig.~\ref{fig:capacity} C).  Since
$\{\textbf{L}_k^{(t)}\}$   are  mutually  orthogonal,   the  different
initial conditions lead to  independent encoding channels.  Again, as
the dynamics are non-normal, the inputs $\textbf{R}_k$ and the outputs
$\textbf{L}_k$  are  not identical,  so  that  the  dynamics for  each
amplified   input  are   at  least   two-dimensional
(Fig.~\ref{fig:capacity} C).

How  many independent,  orthogonal inputs  can a  network  encode with
amplified transients?  To estimate  this number, a central observation
is that  the slopes  of the different  singular value  trajectories at
$t=0$  are given  by  the eigenvalues  of  the symmetric  part of  the
connectivity  $\JS$. This  follows  from the  fact  that the  singular
values  of  the  propagator  $\prop$   are  the  square  root  of  the
eigenvalues  of $\textbf{P}_{t}^T\textbf{P}_{t}$,  and at  short times
$\delta       t$       $\textbf{P}_{\delta      t}^T\textbf{P}_{\delta
  t}\simeq\I+2(\JS-\I)\delta  t$.   This implies  that  the number  of
singular values  with positive slope at  the initial time  is equal to
the  number of  eigenvalues of  the symmetric  part $\JS$  larger than
unity. To  eliminate the  trajectories with very  short amplification,
one  can further  constrain  the slopes  to  be larger  than a  margin
$\epsilon$,  in  which  case  the  number  of  amplified  trajectories
$N_S(\epsilon)$ is given by the  number of eigenvalues of $\JS$ larger
than $1+\epsilon$. Note that $N_S(\epsilon)$ provides only a lower
bound  on the  number of  amplified  inputs, as  singular values  with
initial  slope smaller  than  zero  can increase  at  later times. It is straightforward to compute $N_S(\epsilon)$ when the
connectivity  $\J$   is  Gaussian.   In  this   case  the  probability
distribution of the eigenvalues of its symmetric part $\JS$ follows the
semicircle law (Fig.~\ref{fig:rand}), and  when  the number of neurons $N$
is large, the number $N_s$ of amplified inputs scales linearly with $N$.

To summarize,  the amplified inputs and the  corresponding encoding at
peak amplification can be  determined directly from the singular value
decomposition  of the  propagator,  given by  the  exponential of  the
connectivity  matrix.   For  an  arbitrary $N\times  N$  matrix  $\J$,
characterizing  analytically  the SVD of  its  exponential is  in
general  a complex and  to our  knowledge open  mathematical problem.
For  specific classes  of matrices,  the  propagator and  its SVD  can
however be explicitly computed, and  in the following we will exploit
this approach.

\newcommand{\UV}{\textbf{u}\textbf{v}}
\newcommand{\U}{\textbf{u}}
\newcommand{\V}{\textbf{v}}

\newcommand{\Uu}{\U^{(1)}}
\newcommand{\Vu}{\V^{(1)}}
\newcommand{\Ud}{\U^{(2)}}
\newcommand{\Vd}{\V^{(2)}}

\subsection*{Implementing specific transient trajectories}\

The  approach  outlined above  holds  for  any arbitrary  connectivity
matrix,  and allows  us  to  identify the  external  inputs which  are
strongly amplified  by the recurrent  structure, along with  the modes
that get most activated  during the elicited transients, and therefore
encode the inputs. We now turn to the converse question: how to choose
the network connectivity $\J$  such that it generates a pre-determined
transient trajectory.  Specifically, we  wish to determine the minimal
connectivity  that  transiently transforms  a  fixed, arbitrary  input
$\ic$   into   a   fixed,   arbitrary   output   \textbf{w},   through
two-dimensional dynamics.

\begin{figure}[H]
\centering
\includegraphics[scale=0.95]{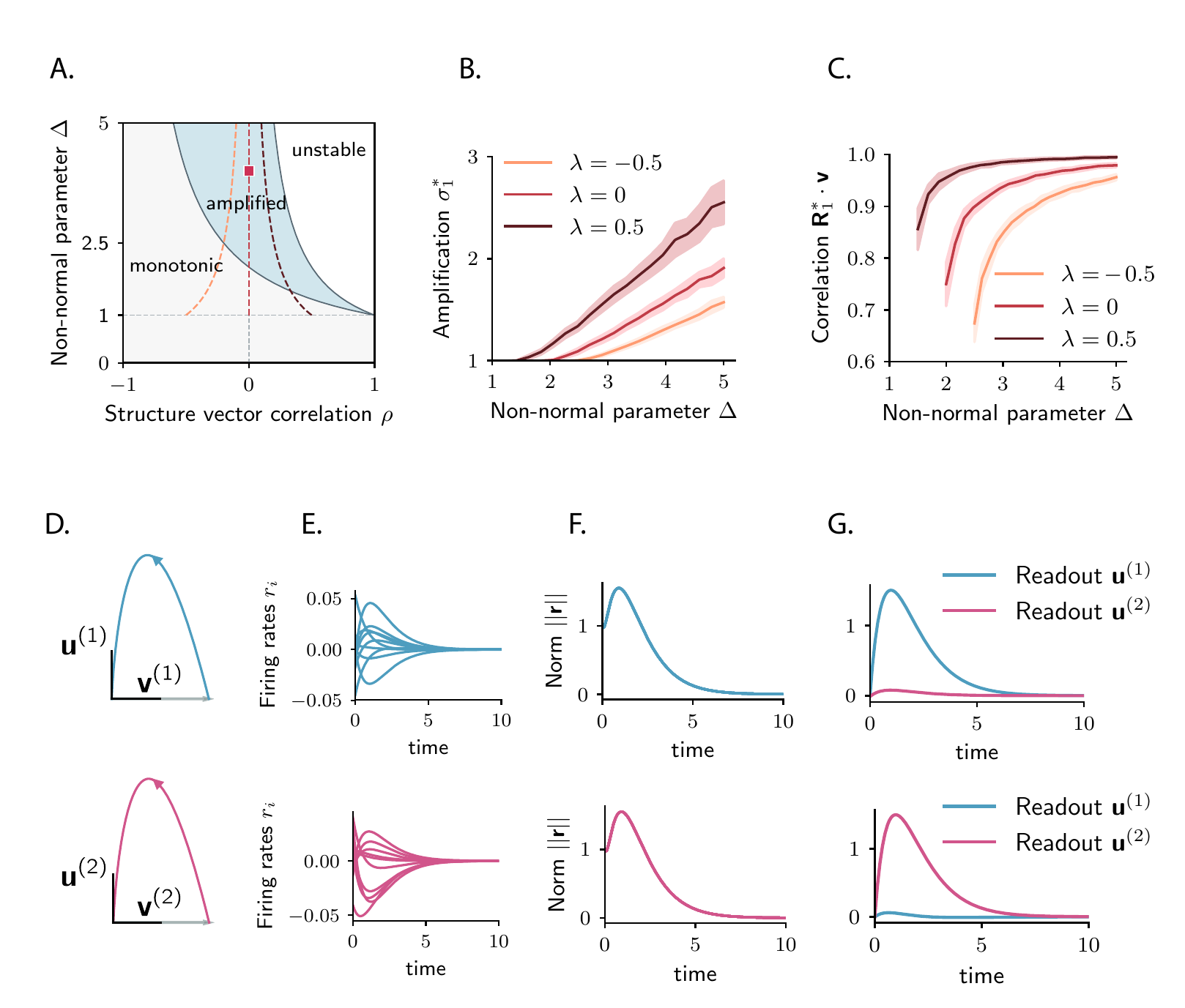} 
\caption{\label{fig:rank-one}
\textsf{\bfseries  Low-dimensional amplified dynamics in random networks with unit-rank structure.} \textsf{\bfseries A.} Dynamical regimes as a function of the structure vector correlation $\rho=\textbf{u}\cdot\textbf{v}$ and the scaling parameter of the connectivity matrix, $\Delta$. Grey shaded areas correspond to parameter regions where the network activity is monotonic for all inputs; blue shaded areas indicate parameter regions where the network activity is amplified for specific inputs; for parameter values in the white area, activity is unstable. Samples of dynamics are shown in the bottom panels, for parameter values indicated by the colored dot in the phase diagram: $\Delta=4$ and  $\rho=0$. Dashed colored traces correspond to the parameter regions explored in panels {\bf B.} and {\bf C.}, defined by the equation $\lambda=\Delta\rho$. \textsf{\bfseries B.} Maximum amplification of the system, quantified by  $\sigma_1(\propp)$, the first singular value of the propagator, as a function of the scaling parameter $\Delta$. Here we fix the eigenvalue of the connectivity matrix  $\lambda=\Delta\rho$ associated with the eigenvector $\textbf{u}$, and vary $\Delta$. Colored traces correspond to different choices of the eigenvalue of the connectivity $\lambda$.  \textsf{\bfseries C.} Correlation between the optimally amplified input direction $\textbf{R}_1^*$ and the structure vector $\textbf{v}$ as a function of the parameter $\Delta$. Increasing the non-normal parameter $\Delta$ aligns the optimally amplified input with the structure vector $\textbf{v}$. In \textsf{\bfseries B.} and \textsf{\bfseries C.} mean and standard deviation over $50$ realizations of the connectivity matrix are shown for each trace. The elements of the structure vectors are drawn from a Gaussian distribution, so that they have on average unit norm and correlation $\rho$ (see \textit{Methods}). \textsf{\bfseries D.} Low-dimensional dynamics in the case of two stored patterns. Input $\V^{(1)}$ (resp. $\V^{(2)}$) elicits a two-dimensional trajectory which brings the activity along the other structure vector $\U^{(1)}$ (resp. $\U^{(2)}$), mapping stimulus $\V^{(1)}$ (resp. $\V^{(2)}$) into its transient readout $\U^{(1)}$ (resp. $\U^{(2)}$). Blue and red colors correspond to the two stored patterns. \textsf{\bfseries E.} Firing rates of 10 individual units. \textsf{\bfseries F.} Temporal evolution of the activity norm . \textsf{\bfseries G.} Projection of the network response evoked by the input along $\V^{(1)}$ (resp. $\V^{(2)}$) on the corresponding readout $\U^{(1)}$ (resp. $\U^{(2)}$). The case of unit rank connectivity (one stored pattern) reduces to the first row of panels ${\textbf{D.}-\textbf{G.}}$ (where the activity on $\U^{(2)}$ is equivalent to the activity on a readout orthogonal to $\U^{(1)}$). $N=3000$ in simulations.
}
\end{figure}

To address this question, we consider a minimal connectivity structure
given       by      a       unit-rank       matrix      $\J=\Delta\uv$
\citep{Hopfield82,Francesca2}.  Here  $\textbf{u}$ and $\textbf{v}$ are
two    vectors   with    unitary   norm    and    correlation   $\rho$
($\langle\textbf{u},\textbf{v}\rangle=\rho$),   and  $\Delta$   is  an
overall  scaling  parameter.   We  applied to  this  connectivity  the
general  analysis outlined  above  (see \textit{Methods}).   The only  non-zero
eigenvalue  of  $\J$ is  $\Delta\rho$,  and  the corresponding  linear
system is  stable for $\Delta\rho<1$.   The largest eigenvalue  of the
symmetric   part    of   the   connectivity   $\JS$    is   given   by
$\Delta(\rho+1)/2$, so that  the network displays amplified transients
if  and only if  $\Delta(\rho+1)/2>1$ (while  $\Delta\rho<1$). Keeping
the  eigenvalue  $\Delta\rho$ constant  and  increasing $\Delta$  will
therefore  lead  to  a   transition  from  monotonically  decaying  to
amplified  transients (Fig.~\ref{fig:rank-one}  A).  If $\rho=0$,  the
vectors  $\textbf{u}$   and  $\textbf{v}$  are   orthogonal,  and  the
condition for amplification is simply $\Delta>2$. Note that in this situation, amplification is obtained without slowing down the dynamics, in contrast to randomly coupled networks \citep{Hennequin2012}.

For  this   unit  rank   connectivity  matrix,  the   full  propagator
$\prop=\exp(t(\J-\I))$  of the  dynamics can  be  explicitly computed
(see \textit{Methods}). The non-trivial dynamics are two-dimensional, and lie
in  the  plane  spanned  by  the structure  vectors  $\textbf{u}$  and
$\textbf{v}$   (Fig.~\ref{fig:rank-one}  B),   while   all  components
orthogonal to this plane  decay exponentially to zero. Determining the
singular value  decomposition of the  propagator allows us  to compute
the  amount  of   amplification  of  the  system,  as   the  value  of
$\sigma_1(\prop)$  at  the  time  of  its maximum  $\tstar$.   In  the
amplified   regime   (for   $\Delta(\rho+1)/2>1$),   the   amount   of
amplification increases  monotonically with $\Delta$.   Since only one
eigenvalue of $\JS$ is larger  than unity, only one input perturbation
is able to generate amplified  dynamics. For large values of $\Delta$,
this optimal input direction is strongly correlated with the structure
vector $\textbf{v}$. Perturbing  along the vector $\textbf{v}$ elicits
a  two-dimensional  trajectory  which  at its  peak  amplification  is
strongly  correlated  with  the  other structure  vector  $\textbf{u}$
(Fig.~\ref{fig:rank-one}  B).  Choosing $\textbf{v}=\textbf{r}_0$  and
$\textbf{u}=\textbf{w}$, the unit-rank connectivity therefore directly
implements a  trajectory that maps  the input $\textbf{r}_0$  into the
output $\textbf{w}$,  identified as  the transient readout  vector for
stimulus $\ic$.

Several,  orthogonal   trajectories  can  be   implemented  by  adding
orthogonal    unit   rank    components.    For    instance,   taking
$\J=\Delta\U^{(1)} \V^{(1)T}+\Delta\U^{(2)}  \V^{(2)T}$, where the planes defined by the structure vectors in each term are mutually orthogonal, the input $\Vu$
    evokes  a trajectory  which is  confined to  the plane  defined by
    $\Uu$ and  $\Vu$, and which maps  the input $\Vu$  into the output
    $\Uu$  at the time  of peak  amplification.  Similarly,  the input
    $\Vd$ is mapped into the  output $\Ud$ during the evoked transient
    dynamics.   Therefore, the  rank-$2$ connectivity  $\J$ implements
    two transient patterns, encoding  the stimuli $\Vu$ and $\Vd$ into
    the readouts $\Uu$ and $\Ud$. A natural question is how robust the
    scheme is and how many patterns can be implemented in a network of
    fixed size $N$.

\subsection*{Robustness and capacity}

\begin{figure}[h]
\centering
\includegraphics[scale=0.8]{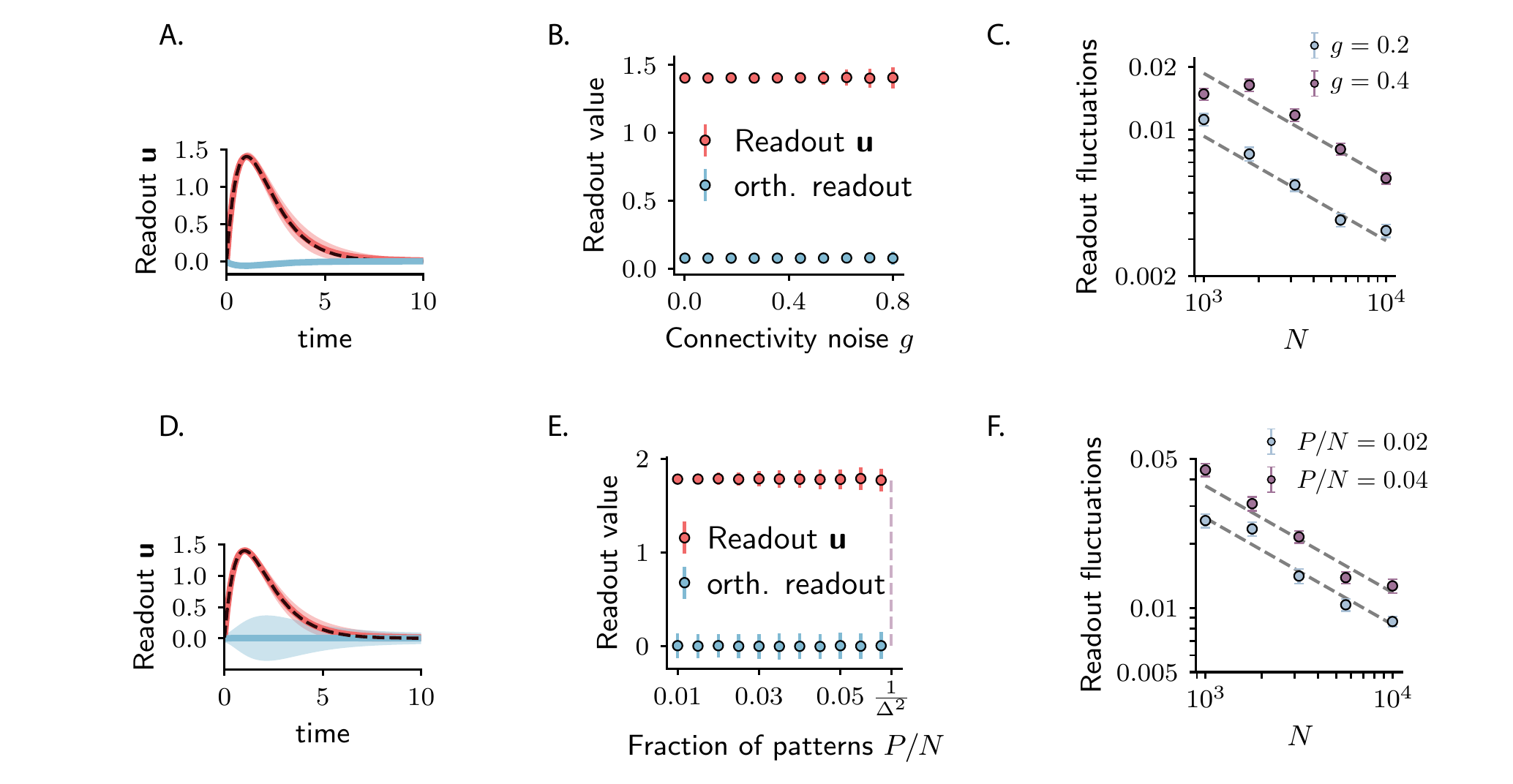} 
\caption{\label{fig-robustness}
\textsf{\bfseries Robustness of the transient coding scheme and capacity of the network}
\textsf{\bfseries (A-B-C)} Robustness of the readout activity for a single stored pattern $\U$-$\V$ in presence of randomness in the connectivity with variance $g^2/N$. \textsf{\bfseries A.} Projection of the population activity elicited by input $\V$ along the readout $\U$ (red trace) and along a readout orthogonal to $\U$ (blue trace) for $g=0.5$. The elements of the orthogonal readout are drawn from a random distribution with mean zero and variance $1/N$ and are fixed over trials. The projection of the activity on $\U$ is also shown for the zero noise case ($g=0$; black dashed line). \textsf{\bfseries B.} Value of the activity along $\U$ (red dots) and along the orthogonal readout (blue dots) at the peak amplification ($t=\tstar$), as a function of $g$. In \textsf{\bfseries A} and \textsf{\bfseries B}, $N=200$; error bars correspond to the standard deviation over $1000$ realizations of the random connectivity. \textsf{\bfseries C.} Standard deviation of the readout activity at the peak amplification as a function of the network size $N$ for two values of $g$. The fluctuations are inversely proportional to the network size and scale as $g\Delta^2/\sqrt{N}$. Error bars correspond to the standard deviation of the mean over 100 realization of the connectivity noise.
\textsf{\bfseries (D-E-F)} Robustness of the transient coding scheme in presence of multiple stored patterns. \textsf{\bfseries D.} Projection of the population activity elicited by one arbitrary amplified input $\V^{(k)}$ along the corresponding readout $\U^{(k)}$ (red trace) and along a different arbitrary readout $\U^{(k^\prime)}$  (blue trace) for $P/N=0.02$. The readout $\U^{(k^\prime)}$ was changed for every trial. The projection of the activity on $\U^{(k)}$ is also shown when only the pattern $\U^{(k)}$-$\V^{(k)}$ is encoded ($P=1$; black dashed line). 
\textsf{\bfseries E.} Value of the activity along $\U^{(k)}$ (red dots) and along the readout $\U^{(k^\prime)}$ (blue dots) at the peak amplification ($t=\tstar$), as a function of $P/N$. In \textsf{\bfseries D} and \textsf{\bfseries E} $N=200$; error bars correspond to the standard deviation over 1000 realizations of the connectivity matrix. \textsf{\bfseries F.} Standard deviation of the readout activity (along $\U^{(k)}$) at the peak amplification as a function of the network size $N$ for two values of $P/N$. The fluctuations are inversely proportional to the network size and scale as $\Delta^3\sqrt{P}/N$. Error bars correspond to the standard deviation of the mean over 100 realizations of the connectivity noise.
}
\end{figure}

To investigate the robustness of the transient coding scheme implemented with unit rank terms, we first examined the effect of additional random components in the connectivity.
 Adding to each connection a random  term of variance $g^2/N$ introduces fluctuations of order $g\Delta^2/\sqrt{N}$ to the component of the activity on the plane defined by $\U$ and $\V$ (see \textit{Methods}). Consequently, the projection of the trajectory on the readout $\textbf{w}=\U$ has fluctuations of the same order (Fig.~\ref{fig-robustness}  A-C). A supplementary  effect of random connectivity is to add to the dynamics a component orthogonal to $\U$ and $\V$, proportional to $\Delta$ (see Appendix H), which however does not contribute to the  readout along $\textbf{w}$. Thus, for large $N$, the randomness in the synaptic connectivity does not impair the decoding of the stimulus $\ic$ from the activity along the corresponding readout $\textbf{w}$.

The robustness of the readouts to random connectivity implies in particular that the unit-rank coding scheme is robust when an extensive number $P$ of orthogonal transient trajectories are implemented by the connectivity $\J$. To show this, we generalize the unit-rank approach and consider a rank-$P$ connectivity matrix, given by the sum of $P$ unit-rank matrices, $\J=\Delta\sum_{p=1}^P \U^{(p)} \V^{(p)T}$, where each term specifies an input-output pair, and all input-output pairs are mutually orthogonal, i.e. uncorrelated. In this situation, the interaction between the dynamics evoked by one arbitrary input $\V^{(p)}$ and the additional $P-1$ patterns is effectively described by a system with connectivity $\J=\Delta\U^{(p)}\V^{(p)T}$ corrupted by a random component with zero mean and variance equal to $\Delta^2P/N^2$ (see \textit{Methods}). From the previous results, it follows that the fluctuations of the activity of the readout $\U^{(p)}$ are of order $\Delta^3\sqrt{P}/N$ (Fig.~\ref{fig-robustness}  D-F). Thus, in high dimension, the readout activity is robust to the interactions between multiple encoded trajectories. When the number of encoded trajectories is extensive ($P=O(N)$), each stimulus $\V^{(p)}$ can therefore still be decoded from the projection of the activity on the corresponding readout $\U^{(p)}$

A natural upper bound on the number of trajectories that can be implemented by the connectivity $\J$ is derived from the stability constraints of the linear system. Indeed, the largest eigenvalues of $\J$ is given by $\Delta\sqrt{P/N}$ and it needs to be smaller than one for stability. Thus, the maximum number of trajectories that can be encoded in the connectivity $\J$ is given by $P_{\rm max}=N/\Delta^2$ and defines the capacity of the network. Crucially, the capacity scales linearly with the size of the network $N$. The capacity also decreases for highly amplified systems, resulting in a trade-off between the separability of the neural activity evoked by different stimuli (quantified by $\Delta$) and the number of stimuli that can be encoded in the connectivity (quantified by $P_{\rm max}$).

\clearpage

\section*{Discussion}\

We examined the conditions under which linear recurrent  networks
can implement an  encoding of stimuli  in  terms  of
amplified transient trajectories. 
The  fundamental mechanism underlying  amplified transients  relies on
the non-normal properties of  the connectivity matrix, i.e. the fact
that the  left- and right-eigenvectors of the  connectivity matrix are
not  identical  \citep{Trefethen578}. A  number  of  recent studies  in
theoretical  neuroscience have pointed  out the  interesting dynamical
properties of networks with non-normal connectivity \citep{White2004,Ganguli2008,
  Murphy2009,      Goldman2009,      Hennequin2012,     Hennequin2014,
  Ahmadian2015, Marti2018}.    Several   of   these  works   \citep{   Murphy2009,
  Hennequin2012,   Hennequin2014,  Ahmadian2015}  have   examined  the
amplification  of  the   norm  of  the  activity  vector,   as  we  do
here. However, it was not pointed out that the presence of amplification can be diagnosed by considering the eigenvalues of the symmetric part $\JS$ of  the connectivity matrix (rather than examining properties of the eigenvectors of the connectivity matrix  $\J$), leading to the distinction of two classes of recurrent networks.  This general criterion appears to be well-known in the theory of linear systems (Theorem 17.1 in \cite{Trefethen578}). Here we applied it to standard models of recurrent networks used in computational neuroscience, and in particular to low-rank networks \citep{Francesca2}.

Applying  the criterion  for transient  amplification  to classical
randomly connected networks, we found that amplification occurs only in a
narrow parameter  region close to the instability,  where the dynamics
substantially slow down  as previously shown \citep{Hennequin2012}. To
circumvent this issue, and produce strong transient amplification away
from     the      instability,     \cite{Hennequin2014}     introduced
stability-optimized circuits (SOCs) in which inhibition is fine-tuned
to closely balance excitation, and demonstrated that such dynamics can
account  for  the  experimental  data  recorded in  the  motor  cortex
\citep{Churchland2012}.   We showed  here that  low-rank  networks can
achieve the same purpose, and  exhibit strong, fast amplification in a
large parameter region away  from the instability. One difference with
SOCs is  that low-rank networks  explicitly implement low-dimensional
dynamics that  transform a specified  initial state into  a specified,
orthogonal output state. Several low-rank channels could be combined to  reproduce higher-dimensional dynamics similar to those observed during the generation of complex movements\citep{Churchland2012}.

The  study by Murphy  and Miller  \citep{Murphy2009} reported  that the
excitatory-inhibitory (EI)  structure  of  cortical  networks 
induces non-normal amplification between so-called  sum and difference E-I modes.
Interestingly,  the  specific  networks  they considered  are  of  the
low-rank type,  with sum and  difference modes corresponding  to left-
and    right-   vectors    of   the    individual    unit-rank   terms
\citep{Ahmadian2015}.   This  connectivity  structure  is  therefore  a
particular   instance  of   the  low-rank   implementation   of  amplified
trajectories  that we  described  here.  Moreover,  Murphy and  Miller
specifically focused on  the inhibition-dominated regime \citep{Ozeki2009},
which as we show approximately corresponds to the class of unit-rank E-I networks
satisfying the general  criterion for transient amplification  (Fig.~\ref{fig:2pop} and Supp Info). In
the  present  study,  we   have  not  enforced  a  separation  between
excitatory  and  inhibitory  neurons,  but  this  can  be  done  in  a
straight-forward way by adding a  unit-rank term in which
all excitatory  (resp.  inhibitory) connections have  the same weight,
and  these weights  are chosen  strong enough  to make  all excitatory
(resp.   inhibitory)   synapses   positive  (resp.   negative).   This
additional component would induce  one more amplified channel that would
correspond to the global E-I difference mode of Murphy and Miller.

Here  our   aim  was  to   produce  amplified,  but   not  necessarily
long-lasting  transients.  The timescale  of the  transients generated
using  the  unit-rank implementation  is  in  fact  determined by  the
effective timescale of the network,  set by the dominant eigenvalue of
the connectivity matrix. As shown  in previous studies that focused on
implementing    transient   memory    traces   \citep{White2004,
  Ganguli2008, Goldman2009}, longer  transients can be  obtained either
by increasing recurrent feedback (i.e. the overlap between vectors in
the   unit-rank  implementation),   or  by   creating   longer  hidden
feed-forward chains.  For instance, an effective feed-forward chain of
length $k$  can be obtained from  a rank $k$ connectivity  term of the
type     $\J=\Delta\V^{(k+1)}     \V^{(k)T}+\ldots     +\Delta\V^{(3)}
\V^{(2)T}+\Delta\V^{(2)}  \V^{(1)T}$, i.e.  in  which each  term feeds
into the next one \citep{Sompolinsky_Kanter86}.  This leads in general to
a $k+1$-dimensional  transient with a  timescale extended by  a factor
$k$ \citep{Goldman2009}.  Implementing this kind of higher-dimensional transients
naturally comes at the cost  of reducing the corresponding capacity of
the network.

The implementation of transient channels proposed here clearly bears a
strong analogy  with Hopfield  networks \citep{Hopfield82}. The  aim of
Hopfield networks is to store  patterns of activity in memory as fixed
points  of  the  dynamics, and  this  is  achieved  by adding  to  the
connectivity matrix a  unit-rank term $\bm{\xi}\bm{\xi}^T$ for
each  pattern  $\bm{\xi}$. One  key  difference  with the  present
network  is  that Hopfield  networks  rely on
symmetric connectivity \citep{Brunel2016}, while  amplified transients are obtained  by using strongly
asymmetric  terms in which  the left-  and right-vectors  are possibly
orthogonal.  Another  difference is that  Hopfield networks rely  on a
non-linearity to generate fixed points for each pattern, while here we
considered  instead  linear  dynamics  in  the vicinity  of  a  single
fixed-point. The non-linearity of  Hopfield networks endows them with
error-correcting  properties,  in  the  sense  that  a  noisy  initial
condition will  always lead  to the activation  of a  single memorized
pattern. A weaker  form of  error-correction is also  present in  our linear,
transient encoding, since any component along non-amplified directions
will  decay faster  than  the  amplified pattern.   However,  if  two
amplified patterns are simultaneously activated, they will lead to the
activation of both corresponding  outputs. This absence of competition
may not be undesirable, as  it can allow for the simultaneous encoding,
and possibly binding, of several complementary stimulus features.

While we  focused here on linear  dynamics in the vicinity  of a fixed
point,  strong non-linearities  can give  rise to  different transient
phenomena \citep{Laje2013}.  In  particular, one prominent proposal is
that  robust   transient  coding  can  be   implemented  using  stable
heteroclinic channels, i.e.  sequences of saddle points that feed into
each other \citep{Rabinovich2008}.   This mechanism has been exploited
in     specific     models     based     on     clustered     networks
\citep{Rabinovich2008_PlosCB}.   A  general theory  for  this type  of
transient  coding   is  to   our  knowledge  currently   lacking,  and
constitutes an interesting avenue for future work.

%%%%%%%%%%%%%%%%%%%%%%%%%%%%%%%%%%%%%%%%%%%%%%%%%
%%%%%%%%%%%%%%%%%%%%%%%%%%%%%%%%%%%%%%%%%%%%%%%%%

\clearpage

\begin{center}
\textsf{\LARGE \bfseries  Methods}
\end{center}

\setcounter{secnumdepth}{-2}
\setcounter{tocdepth}{3}
\textsf{\tableofcontents}

\clearpage

\section{Method details}

\subsection{The network model}

We study a recurrent network of $N$ randomly coupled rate units. Each unit $i$ is described by the time-dependent variable $r_i(t)$, representing its firing rate at time $t$. The transfer function of the individual units is linear, so that the equation governing the temporal dynamics of the network reads:

\begin{equation}\label{rate_model2}
\tau\dot{r}_i=-r_i+\sum_{j=1}^NJ_{ij}r_j+I(t)r_{0,i},
\end{equation}
where $\tau$ represents the membrane time constant (fixed to unity), and $J_{ij}$ is the effective synaptic strength from neuron $j$ to neuron $i$. In absence of external input, the system has only one fixed point corresponding to $r_i=0$ for all $i$. To have stable dynamics, we require that the eigenvalues of the connectivity matrix $\J$ be smaller than unity, i.e. $\mathfrak{Re}\lmax(\J)<1$. We write the external input as the product between a common time-varying component $I(t)$,  and a term $r_{0,i}$ which corresponds to the relative activation of each unit. The terms $r_{0,i}$ can be arranged in a $N$-dimensional vector $\ic$, which we call the external input direction. Here we focus on very short external input durations ($I(t)=\delta(t)$) and on input directions of unit norm ($||\ic||=1$).  This type of input is equivalent to setting the initial condition to $\textbf{r}(0)=\ic$. Since we study a linear system, varying the norm of the input direction would result in a linear scaling of the dynamics.

\subsection{Dynamics of the network}

We first outline the standard approach to the dynamics of the linear network defined by Eq.~\eqref{rate_model2} (see e.g. \citep{DayanAbbott, Strogatz}). The solution of the differential equation given by Eq.~\eqref{rate_model2} can be obtained by diagonalizing the linear system, i.e. by using a change of basis $\textbf{r}=\textbf{V}\tilde{\textbf{r}}$ such that the connectivity matrix in the new basis $\bm{\Lambda}=\textbf{V}^{-1}\J\textbf{V}$ is diagonal. The matrix $\textbf{V}$ contains the eigenvectors $\textbf{v}_1,\textbf{v}_2,..., \textbf{v}_N$ of the connectivity $\J$ as columns, while $\bm{\Lambda}$ has the corresponding eigenvalues $\lambda_i$ on the diagonal. Therefore the variables $\tilde{\textbf{r}}$ represent the components of the rate vector on the basis of eigenvectors of $\J$. In this new basis the system of coupled equations in Eq.~\eqref{rate_model2} reduces to the set of uncoupled equations

\begin{equation}\label{rate_model2_uncoupled}
\dot{\tilde{r}}_i=-\tilde{r}_i+\lambda_i\tilde{r}_i+\delta(t)\tilde{r}_{0,i}.
\end{equation}
The dynamics of the linear network given by Eq.~\eqref{rate_model2} can thus be written in terms of its components on the eigenvectors $\textbf{v}_i$ as 

\begin{equation}
\textbf{r}(t)=\sum_{i=1}^N \tilde{r}_i(t)\textbf{v}_i,\qquad \tilde{r}_i(t)=e^{t(\lambda_i-1)/}\,\tilde{r}_{0,i}.
\end{equation}
Equivalently, the solution of the linear system can be expressed as the product between a linear, time-dependent operator $\prop$ and the initial condition $\ic$ \citep{Arnold}:

\begin{equation}\label{solution_with_propagator}
\textbf{r}(t)=\prop\,\ic.
\end{equation}
The linear operator $\prop$ is called the propagator of the system and it is defined as the matrix exponential of the connectivity matrix $\J$, i.e. $\prop=\exp(t(\J-\I)/\tau)$. By using the definition of matrix exponential in terms of power series, we can express the propagator as ${\prop=\textbf{V}\,\diag(e^{t(\lambda_1-1)},...,e^{t(\lambda_N-1)})\,\textbf{V}^{-1}}$. From Eq.~\eqref{solution_with_propagator} we note that the propagator $\prop$ at time $t$ defines a mapping from the state of the system at time $t=0$, i.e. the external input direction $\ic$, to the state $\textbf{r}(t)$.

\subsection{Dynamics of the norm}

To study the amplification properties of the network, we follow \citep{Caswell97} and  focus on the temporal dynamics of the population activity norm $||\textbf{r}(t)||$ \citep{Hennequin2014}. The equation governing the dynamics of the norm can be derived by writing $||\textbf{r}||=\sqrt{\textbf{r}^T\textbf{r}}$, so that the relative rate of change of the norm is given by \citep{Caswell97}

\begin{equation}
\begin{split}
\frac{1}{||\textbf{r}||}\frac{\deriv ||\textbf{r}||}{\deriv t}&=\frac{1}{\sqrt{\textbf{r}^T\textbf{r}}}\frac{\deriv \sqrt{\textbf{r}^T\textbf{r}}}{\deriv t}\\
&=\frac{1}{2\textbf{r}^T\textbf{r}}
\bigg(  \frac{\deriv \textbf{r}^T}{\deriv t}\textbf{r}+\textbf{r}^T\frac{\deriv \textbf{r}}{\deriv t} \bigg).
\end{split}
\end{equation}
By using Eq.~\eqref{rate_model2} we can write the right hand side of the previous equation as 

\begin{equation}\label{R10}
\begin{split}
\frac{1}{||\textbf{r}||}\frac{\deriv ||\textbf{r}||}{\deriv t}&=
\frac{\textbf{r}^T\big((\J^T-\I)+(\J-\I)\big)\textbf{r}}{2||\textbf{r}||^2}\\
&=\frac{\textbf{r}^T  (\JS-\I)\textbf{r}}{||\textbf{r}||^2},
\end{split}
\end{equation}\
where we introduced $\JS=(\J+\J^T)/2$, the symmetric part of the connectivity matrix $\J$.

Both the eigenvalues and the eigenvectors of $\JS$ provide information on the transient dynamics of the system.
On one hand, we show in the main text that the activity norm can have non-monotonic behaviour if and only if at least one eigenvalue of the matrix $\JS$ is larger than one. Therefore the eigenvalues of $\JS$ determine the type of transient regime of the system. On the other hand, as $\JS$ is symmetric,  its set of eigenvectors is orthogonal and provides a useful orthonormal basis onto which we can project the dynamics. In this basis, the connectivity matrix is given by $\textbf{J}^\prime=\textbf{V}_S^T\J\textbf{V}_S$, where $\textbf{V}_S$ contains the eigenvectors of $\JS$ as columns. The matrix $\J$ can be uniquely decomposed as $\J=\textbf{J}_S+\textbf{J}_A$, where $\textbf{J}_A=(\J-\J^T)/2$ is the anti-symmetric part of $\J$, so that

\begin{equation}\label{symmetric_decomposition}
\textbf{J}^\prime=\diag\big(\lambda_1(\JS),...,\lambda_N(\JS)\big)+\textbf{V}_S^T\textbf{J}_A\textbf{V}_S.
\end{equation}
The first term on the right hand side is a diagonal matrix, while the second term is an anti-symmetric matrix. Since the latter has zero diagonal elements, the new connectivity matrix $\J^\prime$ displays the eigenvalues of $\JS$ on the diagonal. The off-diagonal terms of $\J^\prime$ are given by the elements of $\textbf{V}_S^T\textbf{J}_A\textbf{V}_S$ and represent the strength of the couplings between the eigenvectors of $\JS$. In the amplified regime, some of the eigenvalues of $\JS$ are larger than one, so that without the coupling between the modes of $\JS$, the connectivity $\J^\prime$ would be unstable. However, in our case $\J$ and $\J^\prime$ are stable matrices, meaning that the coupling terms ensure the stability of the overall system. Moreover, varying the strengths of the coupling terms while keeping fixed the diagonal terms affects in a non-trivial way the maximum amplification of the system. Therefore, the decomposition in Eq.~\eqref{symmetric_decomposition} allows us to identify the set of key parameters that controls the maximum amplification of a specific system. In the following, we will systematically use this decomposition  to analyze specific classes of matrices.

\subsection{Amplification}

To identify which inputs are amplified, we examine the dynamics of the activity norm $||\textbf{r}(t)||$ for an arbitrary external input $\ic$. The one-dimensional Eq.~\eqref{R10} alone is not enough to determine the time course of $||\textbf{r}(t)||$, since the right hand side depends on the solution of the $N-$dimensional system Eq.~\eqref{rate_model2}. Therefore, for a specific input $\ic$, we can use Eq.~\eqref{solution_with_propagator} and write the norm of the elicited trajectory as

\begin{equation}
||\textbf{r}(t)||=||\prop\ic||.
\end{equation}

\subsubsection{Input-output mapping between amplified inputs and readouts}

The dynamics elicited in response to an input along an arbitrary direction is in general complex. However, the singular value decomposition (SVD) of the propagator provides a useful way to understand the network dynamics during the transient phase. Any matrix $\textbf{A}$ can be written as

\begin{equation}
\textbf{A}=\textbf{L}\bm{\Sigma}\textbf{R}^T,
\end{equation}
where the matrix $\bm{\Sigma}$ contains the singular values $\sigma_i(\textbf{A})$ on the diagonal, while the columns of $\textbf{L}$ (resp. $\textbf{R}$) are the left (resp. right) singular vectors of $\textbf{A}$, i.e. the eigenvectors of $\textbf{A}\textbf{A}^T$ (resp. $\textbf{A}^T\textbf{A}$). The matrices $\textbf{R}$ and $\textbf{L}$ are unitary, meaning that they separately provide two orthogonal sets of unitary vectors. Thus, we can write the SVD of the propagator as

\begin{equation}\label{SVD_propagator2}
\prop=\sigma_1^{(t)}\textbf{L}_1^{(t)}\textbf{R}_1^{(t)T}+\sigma_2^{(t)}\textbf{L}_2^{(t)}\textbf{R}_2^{(t)T}
+...+\sigma_N^{(t)}\textbf{L}_N^{(t)}\textbf{R}_N^{(t)T}.
\end{equation}
From Eq.~\eqref{SVD_propagator2} we see that, at a given time $t$, the propagator $\prop$ maps each right singular vector $\textbf{R}_k^{(t)}$ into the left singular vector $\textbf{L}_k^{(t)}$, scaled by the singular value $\sigma_k^{(t)}$ (see Eq.~\eqref{SVD_propagator}). Note that for normal systems the singular value decomposition and the eigen-decomposition coincide. In this case the matrices $\textbf{L}$ and $\textbf{R}$ both contain the eigenvectors of $\prop$ as columns, so that $\textbf{L}_k^{(t)}$ and  $\textbf{R}_k^{(t)}$ lie on a single dimension. Instead, for a non-normal system the right and left singular vectors do not align along one direction, and the dynamics of the system in response to an input along $\textbf{R}_k^{(t)}$ spans at least the two dimensions defined by the two vectors $\textbf{R}_k^{(t)}$ and  $\textbf{L}_k^{(t)}$. The vectors $\textbf{R}_k^{(t)}$ for which $\sigma_k^{(t)}>1$ correspond to the amplified inputs at time $t$, while the outputs $\textbf{L}_k^{(t)}$ are the corresponding readouts at time $t$.

\subsubsection{Number of amplified inputs}

The number of amplified inputs at time $t$ is given by the number of singular values $\sigma_k^{(t)}$ larger than unity. To estimate this number, we examine the temporal dynamics of the singular values $\sigma_k^{(t)}$  in time (SV trajectories). We observe that, for a system in the amplified regime ($\lmax(\JS)>1$), at least one of the SV trajectories has non-monotonic dynamics, starting from one at $t=0$ and then increasing before decaying to zero. In fact, the singular values of the propagator at small times $t=\delta t$ are defined as the square roots of the eigenvalues of

\begin{equation}\label{propagator_at_small_times}
\textbf{P}_{\delta t}^T\textbf{P}_{\delta t}=\I+2(\JS-\I)\delta  t+O(\delta t ^2).
\end{equation}
From Eq.~\eqref{propagator_at_small_times} we can compute the singular values of $\textbf{P}_{\delta t}$ as

\begin{equation}
\sigma_k(\textbf{P}_{\delta t})= 1+\big(\lambda_k(\JS)-1\big)\delta t + O(\delta t ^2),
\end{equation}
so that the slope at time $t=0$ of the $k$-th singular value of the propagator is

\begin{equation}\label{slope_singular_values}
\frac{\deriv \sigma_k}{\deriv t}\bigg|_{t=0}=\lambda_k(\JS)-1.
\end{equation}
Eq.~\eqref{slope_singular_values} shows that the number of singular values larger than unity at small times is given by the number of the eigenvalues of $\JS$ larger than unity, which we denote as $N_S$. 

\subsubsection{Maximum amplification of the system}

From Eq.~\eqref{SVD_propagator2} we see that the maximum over initial conditions of the amplification at time $t$ corresponds to the dominant singular value of the propagator, $\sigma_1^{(t)}$. The associated amplified input and corresponding readout are respectively $\textbf{R}_1^{(t)}$ and $\textbf{L}_1^{(t)}$. To obtain the maximum amplification of the system over inputs and over time, we need to compute the time $\tstar$ at which $\sigma_1^{(t)}$ attains its maximum value. Therefore, the value $\sigma_1^{(\tstar)}$ quantifies the maximum amplification over inputs and over time, while $\textbf{R}_1^{(\tstar)}$ and $\textbf{L}_1^{(\tstar)}$ correspond respectively to the most amplified input direction and the associated readout.

Interestingly, it can be shown that the input $\textbf{R}_1^{(\tstar)}\equiv \textbf{R}_1^*$ satisfies the equation (see Appendix A)

\begin{equation}\label{equation_optimal_initial_condition}
\textbf{R}_1^{*T}(\JS-\I)\textbf{R}_1^*=0,
\end{equation}
which depends only on the symmetric part of the connectivity matrix $\JS$. We will exploit this equation to identify the amplified initial condition $\textbf{R}_1^*$ in specific cases. Note that, except for $N=2$, Eq.~\eqref{equation_optimal_initial_condition} does not fully specify the maximally amplified input.

\subsection{Characterizing transient dynamics - summary}

Summarizing, our approach for characterizing its transient dynamics can be divided into three main steps:

\begin{enumerate}
\item Compute $\JS$, along with its eigenvalues and eigenvectors.
\item Compute the propagator of the system $\prop$.
\item Compute the Singular Value Decomposition (SVD) of the propagator.
\end{enumerate}

These three steps can  be in principle performed numerically for any connectivity matrix.
For particular classes of connectivity matrices, we show below that some or all three steps are analytically tractable.

\subsection{Random Gaussian network} \

Here we consider a non-normal random connectivity matrix with synaptic strength independently drawn from a Gaussian distribution

\begin{equation}
J_{ij}\sim \mathcal{N}(0, g^2/N).
\end{equation}
The eigenvalues of $\J$ are complex and uniformly distributed in a circle of radius $g$ \citep{Girko}:

\begin{equation}
P(\lambda)=
\begin{dcases}
\frac{1}{\pi g^2},\quad & |\lambda|\leq g\\
0,\quad &  |\lambda|> g
\end{dcases}
\end{equation}

For this class of matrices, we can analytically determine the condition for amplified transients, and estimate the number of amplified inputs. In the stable regime ($g<1$), the symmetric part of the connectivity $\JS$ can have unstable eigenvalues. In fact, the elements of the symmetric part are distributed according to

\begin{equation}\label{symm_gaussian}
J_{S, ij}\sim
\begin{dcases}
\mathcal{N}(0, g^2/2N),\quad &i\neq j\\
\mathcal{N}(0, g^2/N),\quad &i= j
\end{dcases}
\end{equation}
From random matrix theory we know that the eigenvalues of the matrix given by Eq.~\eqref{symm_gaussian} are real and  distributed according to the semicircle law \citep{Wigner55,Wigner58}:

\begin{equation}
P(\lambda)=
\begin{dcases}
\frac{1}{\pi g^2}\sqrt{2g^2-\lambda^2},\quad & |\lambda|\leq \sqrt{2}g\\
0,\quad &  |\lambda|> \sqrt{2}g
\end{dcases}
\end{equation}
In particular, the spectral radius of $\JS$ is $\sqrt{2}g$ , meaning that $\JS$ has unstable eigenvalues if $1/\sqrt{2}<g<1$.

To estimate the number of amplified initial conditions, we compute the lower bound on their number $N_S(\epsilon)$, i.e. the number of eigenvalues of $\JS$ larger than $1+\epsilon$:

\begin{equation}\label{eq-NS-gaussian}
\begin{split}
\frac{N_S(\epsilon, g)}{N}&=\int_{1+\epsilon}^{\sqrt{2}g} P(\lambda(\JS))\,\deriv \lambda(\JS)\\\\
&=\frac{1}{2}-\frac{1}{2\pi g^2}(1+\epsilon)\sqrt{2g^2-(1+\epsilon)^2}-\frac{1}{\pi}\arctan 
\frac{1+\epsilon}{\sqrt{2g^2-(1+\epsilon)^2}}.
\end{split}
\end{equation}
The number of eigenvalues of $\JS$ is maximum when $g$ is close to (but smaller than) unity. In this case Eq.~\eqref{eq-NS-gaussian} at the first order in $\epsilon$ translates to

\begin{equation}
\begin{split}
\frac{N_S(\epsilon, 1)}{N}=\bigg(\frac{1}{2}-\frac{1}{2\pi}-\frac{1}{\pi}\arctan(1)\bigg)-\frac{1}{2\pi}\epsilon\simeq 0.09-0.16\epsilon.
\end{split}
\end{equation}
Therefore, the maximal capacity of a randomly-connected network is therefore around $10\%$.\\

Computing the SVD of the exponential of a $N$-dimensional random matrix is to our knowledge an open mathematical problem. Therefore, for an arbitrary random connectivity matrix, the maximal amount of amplification and the amplified initial conditions are accessible only by numerically computing the SVD of $\exp(t(\J-\I))$.

\subsection{Two-dimensional system}
In this section we consider connectivity matrices describing networks composed of two interacting units of the form

\begin{equation}\label{connectivity-2x2-model}
\textbf{J}=
\begin{pmatrix}
a & b\\
c & d
\end{pmatrix}.
\end{equation}
The eigenvalues of $\J$ determine the stability of the network and can be expressed in terms of its trace and determinant as follows:

\begin{equation}\label{eq_eigval_J_methods}
\lambda^\pm=\frac{\Tr(\J)\pm\sqrt{\Tr^2(\J)-4\deter(\J)}}{2}.
\end{equation}
For the dynamics to be stable, the largest eigenvalue of $\J$ needs to satisfy $\mathfrak{Re}\lambda^+<1$, equivalent to the requirement that $\Tr(\J)<0$ and $\deter(\J)>0$. Note that if the two eigenvalues $\lambda^\pm$ are real, they are symmetrically centered around $\Tr(\J)/2$ on the real axis; if they are complex conjugates they have real part equal to $\Tr(\J)/2$ and are symmetrically arranged along the imaginary dimension.

\subsubsection{Eigenvalues and eigenvectors of $\JS$}

The condition for transient amplification is determined by the two eigenvalues of $\JS$, which read:

\begin{align}\label{eq_eigval_JS_methods}
\LJSpm&=\frac{\Tr(\J)\pm\sqrt{\Tr^2(\J)-4\deter(\J)+4\Delta^2}}{2},
\end{align}
where we introduced the parameter

\begin{equation}
\Delta=\frac{|b-c|}{2}.
\end{equation}
$\Delta$ represents the difference between the off-diagonal elements of $\J$, and provides a measure of how far from symmetric the connectivity matrix is ($\Delta=0$ meaning symmetric connectivity).
Note that the equation for the eigenvalues of $\JS$ (Eq.~\ref{eq_eigval_JS_methods}) differs from the one for the eigenvalues of $\J$ (Eq.~\ref{eq_eigval_J_methods}) by the additive term $4\Delta^2$ under the square root. 
For $\Delta$ above the critical value

\begin{equation}
\Delta_c=\sqrt{1-\Tr(\J)+\deter(\J)}   
\end{equation}
the rightmost eigenvalue of $\JS$ is larger than one, meaning that specific inputs are transiently amplified. Thus, $\Delta$ is the crucial parameter which determines the dynamical regime of the system.

\subsubsection{Decomposition on the modes of $\JS$}

To identify the parameters which determine the maximum amplification of a system, we project the network dynamics onto the orthonormal basis of eigenvectors of $\JS$. In the new basis the connectivity matrix is given by Eq.~\eqref{symmetric_decomposition}. Interestingly, the non-normal parameter $\Delta$ directly appears in the expression of the anti-symmetric part $\J_{A}$, so that we obtain

 \begin{equation}\label{J_prime}
 \textbf{J}^\prime=
 \begin{pmatrix}
 \LJSp (\Delta)& \Delta\\
 -\Delta & \LJSm (\Delta)
\end{pmatrix}
 \end{equation}
up to a sign of the off-diagonal elements. From Eq.~\eqref{J_prime} we see that the non-normal parameter $\Delta$, which determines the dynamical regime of the system, also represents the strength of the coupling between the modes of $\JS$. For $\Delta>\Delta_c$ we have $\LJSp(\Delta)>1$. Thus, at small times, any component of the dynamics on the first mode of $\JS$ is amplified by an amount proportional to $\LJSp(\Delta)-1$. However, at later times, because of the recurrent feedback of strength $\Delta$ between the modes of $\JS$, the system reaches a finite amount of amplification and relaxes back to the zero fixed point. In the following we examine how the value of $\Delta$ determines the amount of amplification of the system.
 
 \subsubsection{Propagator of the dynamics}
 
To examine the dependence of the maximum amplification of the system on the parameter $\Delta$ we compute the propagator $\prop$ and its SVD. A convenient method to compute the exponential of a matrix is provided in \citep{Leonard_MatrixExponential} (see Appendix B), which we apply to $\J^\prime$ to obtain

\begin{equation}\label{x0}
\exp(t\J^\prime)=x_0(t)\I+x_1(t)\J^\prime,
\end{equation}
where the time-dependent functions $x_0(t)$ and $x_1(t)$ are given by

\begin{subequations}
\begin{align}
&x_0(t)=-\frac{\LJm}{\LJp-\LJm}e^{\LJp t}+\frac{\LJp}{\LJp-\LJm}e^{\LJm t}\\
&x_1(t)=\frac{1}{\LJp-\LJm}e^{\LJp t}-\frac{1}{\LJp-\LJm}e^{\LJm t}.
\end{align}
\end{subequations}
Here $\LJp$ and $\LJm$ are the eigenvalues of $\J$ (Eq.~\ref{eq_eigval_J_methods}).

\subsubsection{SVD of the propagator}

In order to compute the maximum amplification of the system we next  compute the largest singular value of the propagator $\sigma_1(\prop)$ (see Appendix C):

\begin{equation}\label{R15}
\sigma_1(\prop)=e^{-t}\sqrt{E(t)^2+H(t)^2}+e^{-t}\sqrt{F(t)^2},
\end{equation}
where

\begin{equation}\label{R16}
\begin{cases}
&E(t)=x_0(t)+ x_1(t)(\LJSp+\LJSm)/2\\
&F(t)=x_1(t)(\LJSp-\LJSm)/2\\
&H(t)=x_1(t)\Delta
\end{cases}
\end{equation}

\subsubsection{Maximum amplification of the system}

Here we compute the maximal amount of amplification by evaluating the maximum value in time of the amplification envelope $\sigma_1(\prop)$ (Eq.~\ref{R15}), and examine its dependence on the non-normal parameter $\Delta$. In particular we find that, for large values of $\Delta$, this dependence is linear. 

To derive this relationship, we note that the combination $\LJSp-\LJSm=\sqrt{\Tr(\J)^2-4\deter(\J)+4\Delta^2}$ depends on $\Delta$, while $\LJSp+\LJSm=\Tr(\J)$ does not. Therefore in Eq.~\eqref{R16} only the functions $H(t)$ and $F(t)$ depend on $\Delta$. In the amplified regime $\Delta\gg\Delta_c$, we have that $H(t)\gg E(t)$ for times $t\gg 1/\Delta$ (while for small times $\delta\ll 1/\Delta$ we have $E(\delta t)=1+\Tr(\J)\delta t/2\gg \Delta \delta t=H(\delta t)$). In addition, for large values of $\Delta$, we can write $(\LJSp-\LJSm)/2=\Delta+O(\Delta^{-1})$ so that the singular value can be written as

\begin{equation}\label{amplification_2by2_vs_Delta}
\begin{split}
\sigma_1(\prop)&\simeq e^{-t}(|H(t)|+|F(t)|)\simeq \Delta e^{-t}x_1(t),\qquad \text{for}\qquad t\gg 1/\Delta,\,\,\Delta\gg\Delta_c.
\end{split}
\end{equation}
To find the value of the maximum amplification we need to compute the time $\tstar$ of occurrence of the global maximum of $\sigma_1(\prop)$ and the value $\sigma_1(\propp)$. The final result is given by

\begin{align}
&\tstar=\underset{t}{\argmax} \,e^{-t}x_1(t)=\frac{1}{\LJp-\LJm}\log\bigg( \frac{\LJm-1}{\LJp-1}\bigg),\\
&\sigma_1(\propp)=\frac{\Delta}{\LJp-\LJm}\Bigg[ 
\bigg(\frac{\LJm-1}{\LJp-1}\bigg)^{\frac{\LJp}{\LJp-\LJm}}
-\bigg(\frac{\LJm-1}{\LJp-1}\bigg)^{\frac{\LJm}{\LJp-\LJm}}
\Bigg].
\end{align}

The two-dimensional model given by Eq.~\eqref{connectivity-2x2-model} has four free parameters, namely the strengths of the four recurrent connections. In our analysis we fix the values of the trace $\Tr(\J)$ and determinant $\deter(\J)$ of the connectivity matrix, so that the dynamics are stable, and vary the parameter $\Delta$. This implies fixing the eigenvalues $\LJpm$ and the corresponding timescales $\tau^{\pm}=1/(1-\mathfrak{Re}\,\lambda^\pm)$.  This approach allows us to explore how different degrees of symmetry in the connectivity, as quantified by $\Delta$, influence the dynamics while keeping the timescales constant.
Thus, we find that, for $\Delta\gg\Delta_c$, and for fixed $\lambda^\pm$, the maximum amplification of the system scales linearly with the non-normal parameter $\Delta$.

\subsubsection{Optimally amplified initial condition}

Here we compute the optimal input direction $\textbf{R}_1^*$ by solving Eq.~\eqref{equation_optimal_initial_condition}. We parametrize the optimal input by the angle  $\theta^*$ it forms with the first mode of $\JS$, i.e. $\textbf{R}_1^*=(\cos\theta^*, \sin\theta^*)^T$. Thus, Eq.~\eqref{equation_optimal_initial_condition} translates into

\begin{equation}\label{R12}
\LJSp\cos^2\theta^*+\LJSm\sin^2\theta^*-1=0
\end{equation}
which is satisfied by

\begin{equation}\label{angle_oic_2D}
\theta^*=\pm\arctan\sqrt{\frac{\LJSp-1}{1-\LJSm}}.
\end{equation}

\subsection{Rank-1 connectivity}

In this section we consider a unit-rank connectivity matrix defined by

\begin{equation}\label{rank-1_connectivity}
\J=\Delta\uv,
\end{equation}
where the vectors $\textbf{u}$ and $\textbf{v}$ are two $N$-dimensional vectors generated as

\begin{align*}
&\textbf{u}=\textbf{x}_1\\
&\textbf{v}=\rho\, \textbf{x}_1+ \sqrt{1-\rho^2}\,\textbf{x}_2,
\end{align*}
where the vectors $\textbf{x}_1$, $\textbf{x}_2$ and $\textbf{y}$ are $N$-dimensional vectors with components drawn from a Gaussian distribution with mean zero and variance $1/N$ and $\rho$ is a number between $-1$ and $1$ \citep{Francesca2}. The average norm and correlation are given by $\langle \U\cdot\U \rangle=\langle \V\cdot\V \rangle=1$ and $\langle\U\cdot\V\rangle=\rho$, and $\Delta$ is an overall scaling parameter. We consider only positive values of $\Delta$, since a minus sign can be absorbed in the correlation coefficient $\rho$. The matrix $\J$ has $N-1$ eigenvalues equal to zero and one eigenvalue given by $\lambda=\Delta\rho$, associated with the eigenvector $\textbf{u}$. In the two-dimensional plane spanned by $\U$ and $\V$, the direction orthogonal to $\V$ specifies another eigenvector of $\J$ corresponding to one of the zero eigenvalues.

\subsubsection{Eigenvalues and eigenvectors of $\JS$}

We first compute the eigenvalues and eigenvectors of the symmetric part of the connectivity

\begin{equation}
\JS=\Delta\frac{\uv+\textbf{v}\textbf{u}^T}{2}.
\end{equation}
$\JS$ is a rank-2 matrix, meaning it has in general two non-zero eigenvalues given by

\begin{equation}\label{eq_eigenval_JS_rank1_tr_det}
\LJSpm=\frac{\Tr\JS\pm\sqrt{(\Tr\JS)^2-4\deter^\prime\JS}}{2}.
\end{equation} 
Here $\deter^\prime \JS=\LJSp\LJSm$ denotes the determinant of $\JS$ restricted to the $\UV$-plane, i.e. the determinant of the $2\times 2$ matrix $[\U,\U_\perp]^T\JS[\U,\U_\perp]$, where $\U_\perp$ is a vector perpendicular to $\U$ on the $\UV$-plane (the determinant of the full matrix $\JS$ is zero because of the zero eigenvalues of $\JS$). We find that the two non-zero eigenvalues of the symmetric part $\JS$ are given by (see Appendix D)

\begin{equation}\label{eq_eigenval_JS_rank1}
\LJSpm=\frac{\lambda\pm\Delta}{2}.
\end{equation}
Note that the eigenvalues of $\JS$ are symmetrically centered around $\lambda/2$, and their displacement is controlled by the scaling parameter $\Delta$.  The condition for the system to be in the regime of transient amplification is therefore

\begin{equation}\label{condition_amplification_rank1}
\frac{\lambda+\Delta}{2}>1.
\end{equation}

To compute the eigenvectors $\textbf{x}_S^\pm$ associated with the non-zero eigenvalues $\LJSpm$ we have to solve the eigenvector equation

\begin{equation}
(\Delta\uv+\Delta\textbf{v}\textbf{u}^T-2\LJSpm\I)\textbf{x}_S^\pm=0.
\end{equation}
Since the two eigenvectors lie on the $\UV$-plane, we can write them in the form $\textbf{x}_S^+=\textbf{u}+\alpha\textbf{v}$ and $\textbf{x}_S^-=\textbf{u}+\beta\textbf{v}$. Solving the eigenvector equation for $\alpha$ and $\beta$ yields $\alpha=1$ and $\beta=-1$. The two normalized eigenvectors of $\JS$ are thus given by

\begin{equation}
\textbf{x}^\pm_S=\frac{\textbf{u}\pm\textbf{v}}{\sqrt{2(1\pm\rho)}}.
\end{equation}

\subsubsection{Decomposition on the modes of $\JS$}
We can project the dynamics of the system on the basis of eigenvectors of $\JS$.
Let $\textbf{V}_S$ be the $N$-dimensional matrix containing the eigenvectors of $\JS$ as columns:

\begin{equation}
\textbf{V}_S=(\textbf{x}^+_S,\textbf{x}_S^-,\bm{\xi}_1,...,\bm{\xi}_{N-2}),
\end{equation}
where the $\bm{\xi}_i$'s are $N-2$ arbitrary vectors orthogonal to both $\U$ and $\V$. The projection of the connectivity matrix $\J$ onto the modes of $\JS$ yields the new connectivity $\J^\prime$:

\begin{equation}\label{Jprime_rank1}
\begin{split}
\J^\prime=
\textbf{V}_S^T\uv\textbf{V}_S &=
\frac{\Delta}{2}
\begin{pmatrix}
\begin{matrix}
\rho+1 & -\sqrt{1-\rho^2} \\
+\sqrt{1-\rho^2} & \rho-1 
\end{matrix} & \mbox{\large \textbf{0}}\\
 \mbox{\large \textbf{0}} &  \mbox{\large \textbf{0}}
\end{pmatrix}=
\begin{pmatrix}
\begin{matrix}
\LJSp & -\sqrt{\Delta^2-\lambda^2}/2\\
\sqrt{\Delta^2-\lambda^2}/2 & \LJSm
\end{matrix} & \mbox{\large \textbf{0}}\\
 \mbox{\large \textbf{0}} &  \mbox{\large \textbf{0}}
\end{pmatrix}.
\end{split}
\end{equation}
From Eq.~\eqref{Jprime_rank1} we see that the parameter $\Delta$ controls the strength of the coupling between the modes of $\JS$ through the term $\sqrt{\Delta^2-\lambda^2}/2$. Thus, in the following analysis, we examine the amplification properties of the system as a function of the parameter $\Delta$.

\subsubsection{Propagator of the dynamics}

We explicitly compute the expression of the propagator for the unit-rank system. From the definition of matrix exponential in terms of infinite sum of matrix powers we obtain

\begin{equation}
\begin{split}
\exp\big(t\Delta\uv\big)&=\sum_{k=0}^\infty \frac{(t\Delta\uv)^k}{k!}=\I+\frac{\Delta\uv}{\lambda}\big(1+\lambda t+\frac{1}{2}\lambda^2t^2+...-1 \big)\\
&=\I+\Delta\frac{e^{\lambda t}-1}{\lambda}\uv.
\end{split}
\end{equation}
Therefore the final expression for the propagator is given by

\begin{equation}\label{propagator_rank1}
\prop=\exp\big(t(\Delta\uv-\I)\big)=e^{-t}+\Delta e^{-t}\alpha(t,\lambda)\uv,
\end{equation}
where we introduced

\begin{equation}\label{eq_for_alpha}
\alpha(t,\lambda)=\frac{e^{\lambda t}-1}{\lambda}.
\end{equation}
Note that the non-trivial dynamics of the system are restricted to the plane spanned by $\U$ and $\V$. In fact any component of the initial condition orthogonal to this plane decays to zero as $e^{-t}$, as any component orthogonal to $\V$ in the $\UV$-plane. From this it follows that non-monotonic transients occur only if the initial condition of the system has a non-zero component on the structure vector $\V$. 

\subsubsection{SVD of the propagator}

To study how the maximum amplification depends on $\Delta$ we compute the amplification envelope $\sigma_1(\prop)$.
The singular values of the propagator $\prop$ are given by the square roots of the eigenvalues of the matrix $\prop^T\prop$. From Eq.~\eqref{propagator_rank1} we can write

\begin{equation}\label{R13}
\begin{split}
e^{2t}\,\prop^T\prop&=(\I+\Delta\alpha(t,\lambda)\textbf{v}\textbf{u}^T)(\I+\Delta\alpha(t,\lambda)\uv)\\
&=\I+2\Delta\alpha(t,\lambda)\JS+\Delta^2\alpha^2(t,\lambda)\textbf{v}\textbf{v}^T.
\end{split}
\end{equation}
We obtain the expression for the singular values of the propagator $\sigma_{1,2}(\prop)$ as a function of  $\Delta$ and $\lambda$ (see Appendix E):

\begin{equation}\label{sv_prop_rank-1}
2e^{2t}\sigma_{1,2}^2(\prop)=2+2\lambda\alpha(t,\lambda)+\alpha^2(t,\lambda)\Delta^2\pm
\sqrt{\Delta^4\bigg[  \alpha^4(t,\lambda)+\frac{1}{\Delta^2}\Big(4\lambda\alpha^3(t,\lambda)+4\alpha^2(t,\lambda) \Big)   \bigg]}.
\end{equation}
The other $N-2$ singular values of $\prop$ are equal to $e^{-t}$.

\subsubsection{Choice of the free parameters}
For the unit-rank system, two parameters out of $\Delta$, $\lambda$ and $\rho$ can vary independently. Since we set $\Delta$ as a free parameter, we need to fix the second independent parameter. We explore three scenarios, which imply different scalings of $\lambda$ or $\rho$ with the parameter $\Delta$:
 
 \begin{enumerate}
 \item   keep the eigenvalue $\lambda$ constant, so as to fix the timescale $\tau=1/(1-\lambda)$, and vary  $\Delta$.  In this case the correlation $\rho$ between the $\U$ and $\V$ scales according to $\rho=\lambda/\Delta$, meaning that increasing $\Delta$ makes the structure vectors more  orthogonal to each other.

 \item  Fix the correlation between the structure vectors, $\rho$, to a positive value and vary $\Delta$. Increasing $\Delta$ has the effect to increase the timescale of the system $\tau=1/(1-\Delta\rho)$, until a point where the system becomes unstable, i.e.  for $\lambda>1$, or equivalently $\Delta>1/\rho$.
 
 \item  Keep $\rho$ fixed to a negative value. In this case $\Delta$ can be increased without bounds and higher values of $\Delta$ decrease the timescale $\tau$.
 
 \end{enumerate}

\subsubsection{Maximum amplification of the system}
 
The singular values of the propagator given by Eq.~\eqref{sv_prop_rank-1} depend in a complex manner on  $\Delta$ and $\lambda$. To understand how the maximum amplification of the system depends on $\Delta$, we study the limit of very large $\Delta$, defined as 

\begin{equation}\label{strong_amplification_regime}
\Delta\gg2\sqrt{1-\lambda(\Delta)},
\end{equation}
which we call the \textit{strong amplification regime}. Note that in general the eigenvalue $\lambda$ depends on $\Delta$, according to $\lambda(\Delta)=\Delta\rho$. For fixed $\lambda$, Eq.~\eqref{strong_amplification_regime} is given by $\Delta\gg2\sqrt{1-\lambda}$, while for a fixed value of $\rho$, Eq.~\eqref{strong_amplification_regime} translates into $\Delta\gg 2(1-\rho)$ (with the additional constraint $\Delta<1/\rho$ ensuring stability, in case $\rho>0$).
 If condition given by Eq.~\eqref{strong_amplification_regime} is met, we can approximate Eq.~\eqref{sv_prop_rank-1} for times $t\gg 2/\Delta$ as

\begin{equation}
2e^{2t}\sigma^2_1(\prop)\simeq 2+2\alpha(t,\lambda)\lambda+2\alpha^2(t,\lambda)\Delta^2,\qquad t\gg 2/\Delta.
\end{equation}
For large $\Delta$ we can neglect the first two terms on the right hand side and write the largest singular value as

\begin{equation}\label{R2}
\sigma_1(\prop)\simeq \Delta e^{-t}\alpha(t,\lambda),\qquad t\gg 2/\Delta.
\end{equation}
The maximum amplification of the system corresponds to the maximum value in time of $\sigma_1(\prop)$. In the strong amplification regime (Eq.~\ref{strong_amplification_regime}) the time $\tstar$ at which the singular value attains its maximum is independent of $\Delta$ and reads:

\begin{align}\label{tstar_rank1}
\tstar=
\underset{t}{\argmax} \,e^{-t}\alpha(t;\lambda) =\frac{1}{\lambda}\log\frac{1}{1-\lambda}.
\end{align}
Thus, the maximum amplification increases monotonically with $\Delta$:

\begin{align}
\sigma_1(\propp)=g\big(\lambda(\Delta)\big)\,\Delta ,\qquad  g(\lambda)=(1-\lambda)^{\frac{1}{\lambda}-1},
\end{align}
where $g(\lambda)$ is a multiplicative factor which depends on the eigenvalue $\lambda$. Different choices of the free parameters imply different growths of the maximum amplification with $\Delta$:

\begin{enumerate}
\item for $\lambda$ fixed and $\rho=\lambda/\Delta$, the maximum amplification increases linearly with $\Delta$.
\item For $\rho>0$ fixed and $\lambda=\Delta\rho$, the maximum amplification increases monotonically with $\Delta$, until it reaches a value equal to $\Delta$ for $\Delta=1/\rho$ (or $\lambda=1$).
\item For $\rho<0$ fixed and $\lambda=\Delta\rho$, the amplification increases monotonically with $\Delta$, but it saturates at a value given by $1/|\rho|$. This follows from the fact that

\begin{equation}
\lim_{\Delta\rightarrow +\infty}g(\Delta\rho)=\frac{1}{\Delta|\rho|}.
\end{equation}

In the case $\rho=0$ the maximum amplification grows linearly as $\Delta/e$, since

\begin{equation}
\lim_{\rho\rightarrow 0}g(\Delta\rho)=\frac{1}{e}.
\end{equation}
\end{enumerate}

\subsubsection{Optimally amplified initial condition and optimal readout}

Using the result we found for the two dimensional case, Eq.~\eqref{angle_oic_2D} and  Eq.~\eqref{eq_eigenval_JS_rank1}, we can determine the angles $\theta^*_R\equiv\theta(\textbf{R}^*_1)$ and $\theta^*_L\equiv\theta(\textbf{L}^*_1)$ of the optimal initial condition and optimal readout with respect to the first mode of $\JS$ as

\begin{equation}\label{eq_tan_rank1}
\tan\theta^*_{L,R}=\pm\arctan\sqrt{\frac{\LJSp-1}{1-\LJSm}}=\pm\sqrt{\frac{\lambda+\Delta-2}{2-\lambda+\Delta}},
\end{equation}
where the $+$ and $-$ signs correspond respectively to $\theta^*_L$ ans $\theta^*_R$. The optimally amplified initial condition and optimal readout are thus given by

\begin{equation}
\begin{dcases}
&\textbf{R}_1^*=\cos\theta^*_R \textbf{x}^+_S+\sin\theta^*_R \textbf{x}^-_S\\
&\textbf{L}_1^*=\cos\theta^*_L \textbf{x}^+_S+\sin\theta^* _L\textbf{x}^-_S.
\end{dcases}
\end{equation}
Here we examine $\textbf{R}_1^*$ and $\textbf{L}_1^*$  in the strong amplification regime (Eq.~\ref{strong_amplification_regime}). We summarize our results as follows.

\begin{enumerate}
\item For fixed $\lambda$ and $\rho=\lambda/\Delta$, we have

\begin{equation}\label{R4}
\tan\theta^*_R\simeq-1+\frac{2-\lambda}{\Delta}
\end{equation}
up to the first order in $\Delta^{-1}$. In the strong amplification regime the second term on the right hand side is much smaller than unity, so that we can compute $\textbf{R}_1^*$ and $\textbf{L}_1^*$ at the first order in  $\Delta^{-1}$. Denoting by $\textbf{v}^\perp=(\U-\rho\V)/\sqrt{1-\rho ^2}$ and $\textbf{u}^\perp=(\V-\rho\U)/\sqrt{1-\rho ^2}$ respectively the vectors orthogonal to $\V$ and $\U$ in the $\UV$-plane, we can write

\begin{equation}
\begin{dcases}
&\textbf{R}_1^*\propto\V+\frac{1}{2}\frac{2-\lambda}{\Delta}\textbf{v}^\perp\\
&\textbf{L}_1^*\propto\U+\frac{1}{2}\frac{2-\lambda}{\Delta}\textbf{u}^\perp.
\end{dcases}
\end{equation}
In the strong amplification regime the optimal initial condition is thus strongly aligned with $\V$ and the optimal readout with the vector $\U$.

\item For fixed $\rho>0$ and $\lambda=\Delta\rho$, we compute the value of $\tan\theta^*$ for the largest value $\Delta$ can take before the system becomes unstable, i.e. $\Delta=1/\rho$. For this value we have

\begin{equation}
\tan\theta^*_R=-\sqrt{\frac{1-\rho}{1+\rho}}\simeq -1+\rho,\qquad \text{ for } 0<\rho\ll 1.
\end{equation}
Thus we have

\begin{equation}
\begin{dcases}
&\textbf{R}_1^*\propto\V+\frac{\rho}{2}\textbf{v}^\perp\\
&\textbf{L}_1^*\propto\U+\frac{\rho}{2}\textbf{u}^\perp.
\end{dcases}
\end{equation}

\item For fixed $\rho<0$  and $\lambda=\Delta\rho$, we can write

\begin{equation}
\tan\theta^*_R\simeq-1+\bigg(\frac{2}{\Delta}-\rho\bigg),
\end{equation}
so that

\begin{equation}\label{R18}
\begin{dcases}
\textbf{R}_1^*\propto\V+\frac{1}{2}\bigg(\frac{2}{\Delta}-\rho\bigg)\textbf{v}^\perp\\
\textbf{L}_1^*\propto\U+\frac{1}{2}\bigg(\frac{2}{\Delta}-\rho\bigg)\textbf{u}^\perp.
\end{dcases}
\end{equation}
\end{enumerate}
In conclusion we find that, in the strong amplification regime, the optimal input has a strong component on the structure vector $\V$, while the optimal readout is strongly aligned with $\U$. In cases (2) and (3), however, this requires the additional condition that the correlation $\rho$ be small.

\subsection{Robustness of the readout to noise in the connectivity}

\newcommand{\SP}{{s(\textbf{P})}}
\newcommand{\SPi}{{s_i(\textbf{P})}}

 In this section we study the dynamics of the system in presence of noise in the synaptic connectivity. We consider the connectivity matrix given by Eq.~\eqref{rank-1_connectivity}, which implements a single transient pattern, and we add uncorrelated noise of standard deviation $g$ to each weight $\Delta u_i v_j$. The resulting connectivity matrix can be written as the sum of a structured unit-rank part and a Gaussian random matrix of the form \citep{Ahmadian2015}

\begin{equation}\label{rank-1_plus_noise}
\J=\Delta\uv+g\bm{\chi}.
\end{equation}
The elements of $\bm{\chi}$ are independently drawn from a Gaussian distribution with zero mean and variance $1/N$ and are uncorrelated with the structured part. In the limit of large $N$, the matrix $\J$ has one eigenvalue equal to the eigenvalue of the unit-rank part, $\lambda=\Delta\rho$, while the other $N-1$ eigenvalues are uniformly distributed in a circle of radius $g$. This holds under the condition that the operator norm of the unit-rank part $\max_{\textbf{x}}||\Delta\uv\textbf{x}||$ is $O(1)$ \citep{Tao2013}. Since the structure vectors $\U$ and $\V$ have unit norm, the operator norm of the unit-rank part is equal to $\Delta$. Therefore, if $\Delta$ is $O(1)$, the condition for the stability of the system is $\max\{\lambda,g\}<1$.

\subsubsection{Eigenvalues of $\JS$}

To draw the phase diagram of the system, we compute the eigenvalues of the symmetric part of $\J$

\begin{equation}\label{symmetric_part_rank1_plus_noise}
\JS=\Delta\frac{\uv+\textbf{v}\textbf{u}^T}{2}+g\bm{\chi}_S,\qquad \bm{\chi}_S=
\frac{\bm{\chi}+\bm{\chi}^T}{2},
\end{equation}
where $\bm{\chi}_S$ denotes the symmetric part of $\bm{\chi}$. The entries of $\bm{\chi}_S$ are distributed according to

\begin{equation}
\bm{\chi}_{S, ij}\sim
\begin{dcases}
\mathcal{N}(0, 1/2N),\quad &i\neq j\\
\mathcal{N}(0, 1/N),\quad &i= j.
\end{dcases}
\end{equation}
We can express the eigenvalues of $\JS$ as a function of $g$ and of the eigenvalues of the symmetric part of the unit-rank matrix (see Eq.~\ref{eq_eigenval_JS_rank1})  \citep{Benaych2010,Benaych2012}. In particular, the rightmost eigenvalue of $\JS$ is given by

\begin{equation}\label{perturbed_eigsymm}
 \lmax(\JS)=
\begin{dcases}
\frac{\lambda+\Delta}{2}+\frac{g^2}{\lambda+\Delta}, & \text{ if } \lambda+\Delta>\sqrt{2}g\\
 \vphantom{\frac{0}{0}}\sqrt{2}g, & \text{otherwise},
 \end{dcases}
 \end{equation}\
 where $\sqrt{2}g$ corresponds to the spectral radius of $\bm{\chi}_S$. We distinguish two cases:
 
 \begin{enumerate}
 \item if $\sqrt{2}g<1$, $ \lmax(\JS)$ is larger than one only if the two conditions
 
 \begin{equation}
 \begin{dcases}
 \frac{\lambda+\Delta}{2}+\frac{g^2}{\lambda+\Delta}>1\\
 \lambda+\Delta>\sqrt{2}g
 \end{dcases}
 \end{equation}
 are satisfied. The first inequality is satisfied if $\lambda+\Delta<1-\sqrt{1-2g^2}$ or $\lambda+\Delta>1+\sqrt{1-2g^2}$. Since for $\sqrt{2}g<1$ we have $1-\sqrt{1-2g^2}<\sqrt{2}g<1+\sqrt{1-2g^2}$, the condition for the amplified regime becomes 
 
 \begin{equation}\label{R17}
 \lambda+\Delta>1+\sqrt{1-2g^2}.
 \end{equation}
 
 \item If $\sqrt{2}g>1$, the inequality $(\lambda+\Delta)/2+g^2/(\lambda+\Delta)>1$ is always satisfied for $\lambda+\Delta>0$, thus holding also for $\lambda+\Delta>\sqrt{2}g$. From Eq.~\eqref{perturbed_eigsymm} we conclude that, for $\sqrt{2}g>1$,  $\lmax(\JS)$ is larger than one independently of the values of $\lambda$ and $\Delta$.
 
 \end{enumerate}
 In the case $\sqrt{2}g<1$, adding noise in the connectivity has a small effect on the phase diagram of the system. In fact,  Eq.~\eqref{R17} can be approximated as $\lambda+\Delta\gtrsim 2-g^2$, which leads to a correction of order $g^2$ to the condition for the amplified regime in absence of noise (see Fig.~\ref{fig:phase_diagram_noise}).

\newcommand*{\vertbar}{\rule[-1ex]{0.5pt}{4ex}}
\newcommand*{\horzbar}{\rule[.5ex]{4ex}{0.5pt}}

\newcommand{\ru}{r_{\rm u}}
\newcommand{\rv}{r_{\rm v}}

 \subsubsection{Robustness of the readout activity}
 \newcommand{\rPerp}{||\textbf{r}^\perp ||}

Here we examine the magnitude of the fluctuations around the mean activity introduced by the random term in the connectivity given by Eq.~\eqref{rank-1_plus_noise}. In particular we assess the robustness of the readout projection of the response evoked by the optimal stimulus of the noiseless system, i.e. $g=0$ (for a discussion on the effects of the connectivity noise on the activity orthogonal to the $\UV$-plane see Appendix H). For simplicity, we assume that the correlation between the structure vectors, $\rho$, is close to zero, and that the condition for the strong amplification regime is satisfied (Eq.~\eqref{strong_amplification_regime}). Therefore, the optimal stimulus is strongly aligned with $\V$, while the corresponding readout is $\U$. We consider the system

\begin{equation}
\frac{\deriv r_i}{\deriv t}=-r_i+\sum_{j=1}^N\big(\Delta u_iv_j+g\chi_{ij}\big)r_j+\sigma\eta_i(t).
\end{equation}
Each neuron receives independent noise with mean zero, variance $\sigma^2$ and autocorrelation function $\langle\eta_i(t)\eta_j(t^\prime)\rangle=\delta_{ij}\delta(t-t^\prime)$, where the angular brackets represent the average over the noise in the input and in the connectivity.
In the limit of large $N$, the equation for the mean activity depends only on the structured part of the connectivity:

\begin{equation}\label{eq_mean_activity_rank1noise}
\frac{\deriv \langle r_i \rangle}{\deriv t}=-\langle r_i \rangle+\sum_{j=1}^N\Delta u_iv_j \langle r_j \rangle.
\end{equation}
Thus, the mean activity in response to an input along $\V$ is given by (see Eq.~\ref{propagator_rank1})

\begin{align}\label{solution_initial_condition_v}
\langle r_i(t) \rangle&=e^{-t}v_i+\Delta te^{-t}u_i
\end{align}
From Eq.~\eqref{eq_mean_activity_rank1noise} we write the equation for the fluctuations of $r_i(t)$ around the mean activity, $\delta r_i(t)=r_i(t)-\langle r_i(t) \rangle$, as

\begin{equation}\label{eq_fluctuations_rank1noise}
\begin{split}
\frac{\deriv \langle \delta r_i \rangle}{\deriv t}=-\delta r_i+\sum_{j=1}^N\Delta u_iv_j\delta r_j+
\sum_{j=1}^N g\chi_{ij}\langle r_j(t) \rangle+\sigma\eta_i(t),
\end{split}
\end{equation}
where we neglected the corrections to $ \delta r_i$ due to the random component. Using Eq.~\eqref{solution_initial_condition_v} we can write the solution of Eq.~\eqref{eq_fluctuations_rank1noise} as

\begin{equation}
\delta r_i(t)=\sum_{k,l=1}^N\int_{0}^t \Big[e^{(t-s)(\Delta\uv-\I)}\Big]_{ik}\Big(g\chi_{kl}\langle r_l(s) \rangle+\sigma\eta_k(s)\Big)\deriv s.
\end{equation}

The time-dependent correlation matrix $\textbf{C}(t)=\langle \delta \textbf{r}(t)\delta \textbf{r}(t)^T  \rangle$ can be written as the sum of two terms, corresponding to the contributions of the noise in the connectivity (with variance $g^2$) and the noise in the input (with variance $\sigma^2$):

\begin{equation}\label{correlation-matrix}
\begin{split}
C_{ij}(t)&=C_{ij}^g(t)+C_{ij}^\sigma(t)\\
&=\frac{g^2}{N}\sum_{k,l}\int_0^t\int_0^t\deriv s_1\deriv s_2 
[e^{(t-s_1)(\Delta\uv-\I)}]_{ik}[e^{(t-s_2)(\Delta\uv-\I)}]_{jk}
\langle r_l(s_1) \rangle\langle r_l(s_2) \rangle\\
&+\sigma^2\sum_{k=1}^N\int_0^t\deriv s 
[e^{(t-s)(\Delta\uv-\I)}]_{ik}[e^{(t-s)(\Delta\uv-\I)}]_{jk},
\end{split}
\end{equation}
where in the first term in the right hand side we used $\langle\chi_{kl}\chi_{mn}\rangle=\delta_{km}\delta_{ln}/N$. 

We start by computing the first term in Eq.~\eqref{correlation-matrix}. Since the elements of the matrix propagator and the mean activity are known (see Eqs.~\ref{propagator_rank1},~\ref{solution_initial_condition_v}), we can compute $C_{ij}^g(t)$ for a given realization of the structured part (see Appendix G). The variance of the activity along the direction of the readout $\U$ due to the noise in the connectivity is computed by projecting the matrix $\textbf{C}^g$ onto $\U$. In particular we compute the variance of $\delta r_u$ and at the peak of the transient phase ($\tstar\simeq 1$, see Eq.~\eqref{tstar_rank1}).
As a result, the fluctuations of the readout activity at $t=\tstar$ due to the noise in the connectivity read:

\begin{equation}\label{du_g}
\U^T\textbf{C}^g(1)\U=\frac{g^2}{N}e^{-2}\bigg(\frac{\Delta^4}{36}+\frac{\Delta^2}{2}+1\bigg)
\end{equation}
and scale as $g\Delta^2/\sqrt{N}$ (for large $\Delta$).

Computing the variance of the activity along the readout $\U$ due to the input noise yields (see Appendix G)

\begin{equation}\label{du_sigma}
\begin{split}
\U^T\textbf{C}^\sigma(1)\U
=\sigma^2\bigg[\frac{1}{2}-\frac{e^{-2}}{2}+\Delta^2\bigg(\frac{1}{4}-\frac{5}{4}e^{-2}\bigg)\bigg].
\end{split}
\end{equation}
From Eq.~\eqref{correlation-matrix}, we can write the total amount of variability along the readout $\U$ at the peak amplification as

\begin{equation}
\U^T\textbf{C}(1)\U=\frac{g^2}{N}e^{-2}\bigg(\frac{\Delta^4}{36}+\frac{\Delta^2}{2}+1\bigg)+
\sigma^2\bigg[\frac{1}{2}-\frac{e^{-2}}{2}+\Delta^2\bigg(\frac{1}{4}-\frac{5}{4}e^{-2}\bigg)\bigg].
\end{equation} 
Note that the fluctuations along $\U$ due to the noise in the input do not depend on the size of the network $N$. 
Therefore, in the limit of large $N$, only the input noise affects the readout activity significantly. By computing the signal-to-noise ratio (SNR) of the readout activity, we can assess the reliability of the readout in presence of input noise. The signal of the readout is simply the amplification level at the peak of the transient phase. Since for orthogonal structure vectors ($\rho\simeq 0$) the amplification grows as $\Delta/e$, we find

\begin{equation}\label{SNR}
SNR(\sigma; \Delta)=\frac{\Delta}{e\,\sigma\,\sqrt{\bigg[\dfrac{1}{2}-\dfrac{e^{-2}}{2}+\Delta^2\bigg(\dfrac{1}{4}-\dfrac{5}{4}e^{-2}\bigg)\bigg]}}.
\end{equation}
The readout is reliable if its signal-to-noise ratio is much larger than unity. 
Interestingly, for large values of $\Delta$ (see Eq.~\ref{strong_amplification_regime}), the SNR is independent of $\Delta$, so that increasing the amplification does not improve the SNR significantly (see Fig.~\ref{fig:SNR}).
In fact, for $\Delta\gg 2$, we can approximate Eq.~\eqref{SNR} as

\begin{equation}
SNR(\sigma; \Delta\gg 2)=\frac{1}{e\,\sigma\,\sqrt{\dfrac{1}{4}-\dfrac{5}{4}e^{-2}}}.
\end{equation}
In this regime, the critical value of $\sigma$ above which the SNR becomes smaller than unity is:

\begin{equation}
\sigma_c=\frac{1}{e\,\sqrt{\dfrac{1}{4}-\dfrac{5}{4}e^{-2}}}\simeq 1.17.
\end{equation}

\subsection{Robustness to multiple stored patterns and capacity of the network}

In this section we examine the robustness of the transient readouts when $P$ transient trajectories are encoded in the connectivity $\J$. We consider a connectivity matrix given by the sum of $P$ unit-rank matrices

\begin{equation}\label{J_P_patterns}
\J=\Delta\sum_{p=1}^P \U^{(p)} \V^{(p)T},
\end{equation}
where the elements of the vectors $\U^{(p)}$ and $\V^{(p)}$ are randomly distributed with zero mean and variance equal to $1/N$. Therefore, for large $N$ and for $P\leq N/2$, these vectors are close to orthogonal to each other, meaning that the correlation between all the pairs of structure vectors, $\rho$, is close to zero. For simplicity, we assume that the non-normal parameter $\Delta$ is the same for all stored trajectories. We first study the case of two stored transient trajectories ($P=2$), then generalizing to an extensive number of patterns $P=O(N)$.

\subsubsection{Two encoded transient trajectories}

The connectivity matrix in this case is given by

\begin{equation}\label{J_2_patterns}
\J=\Delta\U^{(1)}\V^{(1)T}+\Delta\U^{(2)}\V^{(2)T}.
\end{equation}
Since the four structure vectors in Eq.~\eqref{J_2_patterns} are uncorrelated with each other, in the limit of large $N$, we can factorize the full propagator of the dynamics as the product of the propagators of the single unit-rank parts (see Eq.~\eqref{propagator_rank1}) and obtain (see Appendix I)

\begin{equation}\label{propagator_2_patterns}
\begin{split}
\exp\big( t(\J-\I)  \big)&\simeq
e^{-t}\exp(t\Delta\U^{(1)}\V^{(1)T})\exp(t\Delta\U^{(2)}\V^{(2)T})\\
&=e^{-t}\big(\I+\Delta\alpha(t;0)\U^{(1)}\V^{(1)T} \big)\big(\I+\Delta\alpha(t;0)\U^{(2)}\V^{(2)T} \big),
\end{split}
\end{equation} 
where $\alpha(t;\lambda=0)=t$ (see Eq.~\ref{eq_for_alpha}). From Eq.~\eqref{propagator_2_patterns} we see that, in high dimensionality, the two transient patterns do not interact. In fact, any initial condition defined on the plane spanned by $\Uu$ and $\Vu$ evokes a two-dimensional trajectory which remains confined on the same plane. The same holds for the dynamics on the plane defined by $\Ud$ and $\Vd$.  

\subsubsection{Extensive number of encoded trajectories and capacity of the network}

When the number of encoded trajectories $P$ is of order $N$, we cannot factorize the propagator as in the case of two stored patterns, due to the stronger correlations between the $2P$ structure vectors $\U^{(p)}$ and $\V^{(p)}$. However, the results for the case of one stored pattern with connectivity noise can be applied to this case if we write the connectivity matrix in Eq.~\eqref{J_P_patterns} as

\begin{equation}\label{J_P_patterns_split}
\J=\Delta\Uu\V^{(1)T}+\Delta\sum_{p=2}^P \U^{(p)}\V^{(p)T}.
\end{equation}
Here we isolate the first term of the sum but, since all the $P$ patterns are statistically equivalent, the choice of the first pattern is arbitrary. The vectors $\U^{(i)}$ and $\V^{(i)}$ are uncorrelated with each other, so that we can consider the second term on the right hand side of Eq.~\eqref{J_P_patterns_split} effectively as noise in the connectivity $\J=\Delta\Uu\V^{(1)T}$, with mean zero and variance $\Delta^2P/N^2$. In fact, the mean and the variance of the effective noise are given respectively by

\begin{equation}
\sum_{p=2}^P\langle \U^{(p)}_i\V^{(p)}_j\rangle=\sum_{p=2}^P\langle \U^{(p)}_i\rangle\langle \V^{(p)}_j\rangle=0
\end{equation}
and

\begin{equation}
\begin{split}
\sum_{p,q=2}^P \langle \U^{(p)}_i\V^{(p)}_j\U^{(q)}_i\V^{(q)}_j\rangle=
&\sum_{p,q=2}^P \langle \U^{(p)}_i\U^{(q)}_i\rangle\langle\V^{(p)}_j\V^{(q)}_j\rangle=\sum_{p,q=2}^P \frac{1}{N^2}\delta_{pq}\simeq \frac{P}{N^2}.
\end{split}
\end{equation}
Applying the results from the previous sections with $g=\Delta\sqrt{P/N}$, we can state that the noise coming from the additional $P-1$ patterns adds fluctuations of the order $\Delta^3\sqrt{P}/N$ to the projection of the activity on the readout $\Uu$ corresponding to the stimulus $\Vu$. Since the number of encoded patterns $P$ is extensive, the readout fluctuations scale as $1/\sqrt{N}$.

However, when a number $P$ of trajectories are encoded in $\J$, we are not guaranteed that the connectivity has stable eigenvalues. Indeed, the eigenvalues of the matrix $\Delta\sum_{p=2}^P \U^{(p)}\V^{(p)T}$ are distributed in a circle of radius $g=\Delta\sqrt{P/N}$ (yet the spectral density is not uniform, since Eq.~\eqref{J_P_patterns} can be written as the product of two rectangular  Gaussian matrices) \citep{Burda2011}. Thus, to ensure overall stability we need $g=\Delta\sqrt{P/N}<1$, resulting in a maximal number of patterns $P_{\rm max}$ that can be stored in the connectivity before the system becomes unstable. This number defines the capacity of the system and is given by

\begin{equation}\label{eq_capacity}
P_{\rm max}=\frac{1}{\Delta^2}\,N.
\end{equation}
From Eq.~\eqref{eq_capacity} we see that, for fixed $\Delta$, the number of transient trajectories that we can encode in the connectivity matrix scales linearly with the size of the system, $N$. The capacity of the system rapidly drops when $\Delta$ is increased, meaning that more amplified systems can encode less number of stimuli.
When the structure vectors are orthogonal to each other as in our case ($\rho\simeq  0$), the system is amplified for $\Delta>2$ (see Eq.~\ref{condition_amplification_rank1}). Therefore, Eq.~\eqref{eq_capacity} evaluated at $\Delta=2$ provides an upper bound on the capacity for an amplified system with uncorrelated structure vectors:

\begin{equation}
P_{\rm max}<0.25\,N.
\end{equation}

\clearpage
\section{Supplementary information}
\subsection{Appendix A}

 By the definition of singular value decomposition we can express the $k$-th singular value of $\prop$ as $\sigma_k=\textbf{L}_k^T\prop\textbf{R}_k$ and the squared $k$-th singular values as $\sigma_k^2=\textbf{L}_k^T\prop\prop^T\textbf{L}_k$. By differentiating $\sigma^2_k$ we can write

\begin{align}\label{eq_evolution_sigmak}
\frac{\dd \sigma^2_k}{\dd t}=
\dot{\textbf{L}}_k^T\prop\prop^T\textbf{L}_k+
\textbf{L}_k^T\dot{\textbf{P}}\prop^T\textbf{L}_k+
\textbf{L}_k^T\prop\dot{\textbf{P}}^T\textbf{L}_k+
\textbf{L}_k^T\prop\prop^T\dot{\textbf{L}}_k
\end{align}
The first term can be expressed as $\dot{\textbf{L}}_k^T\prop\prop^T\textbf{L}_k=\dot{\textbf{L}}_k^T\textbf{L}_k\sigma_k^2$. Since the right singular vector $\textbf{L}_k$ has unit norm, the scalar product between $\textbf{L}_k$ and its derivative $\dot{\textbf{L}}_k$ is equal to zero. The same holds for the last term on the right hand side $\textbf{L}_k^T\prop\prop^T\dot{\textbf{L}}_k$. Thus, we can rewrite Eq.~\eqref{eq_evolution_sigmak} as

\begin{equation}\label{deriv_sing_val}
\begin{split}
\frac{\dd \sigma^2_k}{\dd t}&=\textbf{L}_k^T\prop(\J+\J^T-2\I)\prop^T\textbf{L}_k\\
&=2\sigma^2_k\,\textbf{R}_k^T(\JS-\I)\textbf{R}_k
\end{split}
\end{equation}
where the last equality follows from $\textbf{L}_k^T\prop=\textbf{L}_k^T\sum_j \sigma_j\textbf{L}_j\textbf{R}^T_j=\sigma_k\textbf{R}^T_k$. By definition, at the optimal time $\tstar$, the derivative of the largest singular value $\sigma_1(\prop)$ vanishes. Since $\sigma_1^2$ is a monotonic function of $\sigma_1$, at time $\tstar$ also the derivative in Eq.~\eqref{deriv_sing_val} vanishes. Thus, the optimal initial condition $\textbf{R}_1^{(\tstar)}$ satisfies  Eq.~\eqref{equation_optimal_initial_condition}.

Following the same steps we can obtain the same equation in terms of the left singular values of the propagator $\textbf{L}_k$:

\begin{equation}
\frac{\dd \sigma^2_k}{\dd t}=
2\sigma^2_k\,\textbf{L}_k^T(\JS-\I)\textbf{L}_k
\end{equation}

\subsection{Appendix B}

For any N-dimensional matrix $\textbf{A}$, we can express its exponential as

\begin{equation}
\exp(t\textbf{A})=\sum_{j=0}^{N-1}\,x_j(t)\textbf{A}^j,
\end{equation}
where the $x_j(t)$ ($0\leq j\leq N-1$) are the $N$ solutions of the $N$-th order differential equation

\begin{equation}\label{Leonard_differential_equation}
x^{(N)}+c_{N-1}x^{(N-1)}+...+c_{1}x^{(1)}+c_0x=0
\end{equation}
with the set of $N$ initial conditions $x_j^{(l)}(0)=\delta_{jl}$ \citep{Leonard_MatrixExponential}

\begin{equation}
\begin{dcases}
x_0^{(0)}=1\\
x_0^{(1)}=0\\
\vdots\\
x_0^{(N-1)}=0
\end{dcases}
\,
,
\qquad
\begin{dcases}
x_1^{(0)}=0\\
x_1^{(1)}=1\\
\vdots\\
x_1^{(N-1)}=0
\end{dcases}
\,
,
\quad
\cdots
\quad
\begin{dcases}
x_{N-1}^{(0)}=0\\
x_{N-1}^{(1)}=0\\
\vdots\\
x_{N-1}^{(N-1)}=1
\end{dcases}
\end{equation}
with $0\leq j\leq N-1$ and $0\leq l\leq N-1$. $x_j^{(n)}$ denotes the $n$-th derivative of the solution $x_j(t)$, while the numbers $c_i$ are the coefficients in the expression of the characteristic polynomial of $\textbf{A}$

\begin{equation}
\deter(\lambda\I-\textbf{A})=\lambda^N+c_{N-1}\lambda^{N-1}+...+c_1\lambda+c_0.
\end{equation}

\subsection{Appendix C}

We express the eigenvalues of $\JS$ as

\begin{equation}
\begin{dcases}
\LJSp=\dfrac{\LJSp+\LJSm}{2}+\dfrac{\LJSp-\LJSm}{2}\\
\LJSm=\dfrac{\LJSp+\LJSm}{2}-\dfrac{\LJSp-\LJSm}{2}
\end{dcases}
\end{equation}
and split the propagator into the sum of two terms:

\begin{equation}\label{convenient_form_prop_2D}
\begin{split}
\exp(t\J^\prime)
&=\begin{pmatrix}
x_0(t)+\frac{\LJSp+\LJSm}{2} \,x_1(t)& x_1(t)\Delta\\
-x_1(t)\Delta & x_0(t)+\frac{\LJSp+\LJSm}{2}\, x_1(t)
\end{pmatrix}+
\begin{pmatrix}
\frac{\LJSp-\LJSm}{2}\,x_1(t) & 0\\
0 & -\frac{\LJSp-\LJSm}{2}\,x_1(t)
\end{pmatrix}\\\\
&=\begin{pmatrix}
E(t) & H(t)\\
-H(t) & E(t)
\end{pmatrix}+
\begin{pmatrix}
F(t) & G(t)\\
G(t) & -F(t)
\end{pmatrix},
\end{split}
\end{equation}
where the time-dependent functions $E,F,G,H$ are given by

\begin{equation}\label{R16-2}
\begin{cases}
&E(t)=x_0(t)+ x_1(t)(\LJSp+\LJSm)/2\\
&F(t)=x_1(t)(\LJSp-\LJSm)/2\\
&G(t)=0\\
&H(t)=x_1(t)\Delta.
\end{cases}
\end{equation}
If we write the SVD of $\prop^\prime$ as

\begin{equation}
e^{t(\J^\prime-\I)}=\textbf{L}\bm{\Sigma}\textbf{R}^T=
\begin{pmatrix}
\cos\beta & \sin\beta\\
-\sin\beta & \cos\beta
\end{pmatrix}
\begin{pmatrix}
\sigma_1 & 0\\
0 & \sigma_2
\end{pmatrix}
\begin{pmatrix}
\cos\gamma & \sin\gamma\\
-\sin\gamma & \cos\gamma
\end{pmatrix}
\end{equation}\\
we can express the time-dependent parameters $\sigma_1(t), \sigma_2(t), \beta(t), \gamma(t)$  ($\sigma_1\geq\sigma_2$) as functions of $E(t), F(t), G(t), H(t)$ (Eq.~\eqref{R16-2}):

\begin{equation}\label{parsSVD}
\begin{split}
&\sigma_1(t)=e^{-t}\sqrt{E^2+H^2}+e^{-t}\sqrt{F^2+G^2}\\
&\sigma_2(t)=e^{-t}\sqrt{E^2+H^2}-e^{-t}\sqrt{F^2+G^2}\\
&2\gamma(t)={\rm atan}(H/E)+{\rm atan}(G/F)\\
&2\beta(t)={\rm atan}(H/E)-{\rm atan}(G/F).
\end{split}
\end{equation}

\subsection{Appendix D}

Since the trace is a linear operator and the trace of $\J$ is equal to the trace of $\J^T$, we can express the trace of $\JS$ as

\begin{equation}\label{trace_JS_rank1}
\Tr\JS=\frac{\Tr(\J+\J^T)}{2}=\LJSp+\LJSm=\lambda.
\end{equation}
The determinant $\deter^\prime\JS$ is simply given by the product of the eigenvalues of $\JS$:

\begin{equation}\label{constrained_det_rank1}
2\deter^\prime\JS=2\LJSp\LJSm=(\LJSp+\LJSm)^2-{\LJSp}^2-{\LJSm}^2=(\Tr\JS)^2-\Tr(\JS^2).
\end{equation}
The last equality in Eq.~\eqref{constrained_det_rank1} follows from the fact that the trace of the square of a matrix is the sum of its squared eigenvalues. Computing $\JS^2$ yields

\begin{equation}
\begin{split}
4\JS^2=&\Delta^2(\uv+\textbf{v}\textbf{u}^T)(\uv+\textbf{v}\textbf{u}^T)= 2\lambda\JS+\Delta^2\textbf{u}\textbf{u}^T+\Delta^2\textbf{v}\textbf{v}^T.
\end{split}
\end{equation}
Thus $\Tr(\JS^2)=(\lambda^2+\Delta^2)/2$ and $2\deter^\prime\JS=(\lambda^2-\Delta^2)/2$. It follows that the eigenvalues of $\JS$ (Eq.~\ref{eq_eigenval_JS_rank1_tr_det}) are given by $\LJSpm=(\lambda\pm\Delta)/2$.

\subsection{Appendix E}

To compute the singular values of the propagator for the unit-rank system, it is convenient to express the matrix 

\begin{equation}
\begin{split}
e^{2t}\,\prop^T\prop&=\I+2\alpha(t;\lambda)\JS+\Delta^2\alpha^2(t;\lambda)\textbf{v}\textbf{v}^T
\end{split}
\end{equation}
in the basis of the eigenvectors of $\JS$. While the second term on the right hand side yields a diagonal contribution proportional to $\diag(\LJSp,\LJSm)$, for the third term we obtain

\begin{equation}
\textbf{V}_S^T\textbf{v}\textbf{v}^T\textbf{V}_S =
\frac{1}{2}
\begin{pmatrix}
\rho+1 & -\sqrt{1-\rho^2}\\
-\sqrt{1-\rho^2} & 1-\rho
\end{pmatrix}.
\end{equation}
The squared singular values of the propagator $\prop$ are therefore the eigenvalues of the matrix

\begin{equation}\label{Matrix1}
e^{2t}\,\prop^T\prop=\begin{pmatrix}
(\rho+1)(a+b)+1 & -b\sqrt{1-\rho^2}\\
-b\sqrt{1-\rho^2} & (\rho-1)(a-b)+1
\end{pmatrix},
\end{equation}
where we defined $a=\Delta\alpha(t;\lambda)$ and $2b=\Delta^2\alpha^2(t;\lambda)$. Thus, we have

\begin{equation}\label{eigvals_propprop_rank-1}
e^{2t}\sigma_{1,2}^2(\prop)= 1+a\rho+b\pm\sqrt{a^2+b^2+2ab\rho},
\end{equation}
where $\sigma_1(\prop)>\sigma_2(\prop)$. Therefore, expanding Eq.~\eqref{eigvals_propprop_rank-1} we obtain the two singular values $\sigma^\pm$ of $\prop$:

\begin{equation}
\begin{split}
e^{2t}\sigma_{1,2}^2(\prop)&=1+\Delta\alpha(t)\rho+\frac{\Delta^2\alpha(t)^2}{2}\pm
\Delta\alpha(t)\sqrt{\Delta\alpha(t)\rho+\frac{\Delta^2\alpha(t)^2}{4}+ 1}.
\end{split}
\end{equation}

\subsection{Appendix F}

An interesting application of the results that we found for the unit-rank connectivity is the system composed of one excitatory and one inhibitory populations. The interactions between the two populations are described by the connectivity matrix

\begin{equation}\label{matrix_2x2_EI}
\J=
\begin{pmatrix}
w & -kw\\
w & -kw
\end{pmatrix},
\end{equation}
where $w$ is the excitatory weight and $k$ represents the relative strength of inhibition with respect to the strength of excitation. We consider the regime in which inhibition is stronger than excitation, i.e. $k>1$. Since $\J$ has unit rank, we can express it in the form given by Eq.~\eqref{rank-1_connectivity}, where

\begin{equation}
\begin{dcases}
\Delta=w\sqrt{2(1+k^2)}\\
\U=\dfrac{(1,1)^T}{\sqrt{2}}\\
\V=\dfrac{(1,-k)^T}{\sqrt{1+k^2}}.
\end{dcases}
\end{equation}
Therefore, $\J$ has only one eigenvalue equal to

\begin{equation}
\lambda=w(1-k)
\end{equation}
and one zero eigenvalue, while the correlation between the structure vectors $\U$ and $\V$ is given by

\begin{equation}
\rho=\frac{1-k}{\sqrt{2(1+k^2)}}.
\end{equation}
Note that the correlation $\rho$ depends only on $k$. For simplicity, we assume that $k$ is fixed and slightly larger than unity:

\begin{equation}\label{approximated_k}
k=1+\epsilon_k.
\end{equation}
Thus, for $\epsilon_k\ll 1$, the parameters of the network are given by

\begin{equation}
\begin{dcases}
\Delta=2w\bigg(1+\dfrac{\epsilon_k}{2}\bigg)\\
\U=\dfrac{(1,1)^T}{\sqrt{2}}\\
\V=
\frac{1}{\sqrt{2}}
\begin{pmatrix}
1\\
-1
\end{pmatrix}
-\dfrac{\epsilon_k}{2}\U\\
\rho=-\dfrac{\epsilon_k}{2}.
\end{dcases}
\end{equation}

Computing the symmetric part of the connectivity $\J$ yields

\begin{equation}
\JS=
\begin{pmatrix}
w & w(1-k)/2\\
w(1-k)/2 & -kw\\
\end{pmatrix},
\end{equation}
which has eigenvalues

\begin{equation}
\LJSpm=\frac{w}{2}(1-k)\pm \frac{w}{2}\sqrt{2(1+k^2)}.
\end{equation}
The condition  $\LJSp(w,k)>1$ determines the region of the parameters $w$ and $k$ where transient amplification occurs. Interestingly, in the inhibition-dominated regime, our approach recovers the results from \citep{Murphy2009}, showing that the system is amplified if the excitatory strength $w$ is (approximately) larger than one. In fact, if $k$ is given by Eq.~\eqref{approximated_k}, we can write

\begin{equation}
\LJSp\simeq w\bigg(1+\frac{\epsilon_k^2}{4}\bigg)
\end{equation}
so that the transient regime is defined by the condition

\begin{equation}
w\gtrsim 1-\frac{\epsilon_k^2}{4}
\end{equation}

In the regime of strong amplification (Eq.~\ref{strong_amplification_regime}), we can compute the optimal initial condition $\textbf{R}^*_1$ and the corresponding readout vector $\textbf{L}^*_1$. If Eq.~\eqref{approximated_k} holds, the strong amplification condition is simply given by $w\gg 1$. Using Eq.~\eqref{R18} we find

\begin{equation}
\begin{dcases}
\textbf{R}^*_1=\frac{1}{\sqrt{2}}
\begin{pmatrix}
1\\
-1
\end{pmatrix}
+\frac{1}{\sqrt{2}}
\bigg(\frac{1}{2w}-\frac{\epsilon_k}{4}\bigg)
\begin{pmatrix}
1\\
1
\end{pmatrix}\\
\textbf{L}^*_1=\frac{1}{\sqrt{2}}
\begin{pmatrix}
1\\
1
\end{pmatrix}
+\frac{1}{\sqrt{2}}
\bigg(\frac{1}{2w}+\frac{\epsilon_k}{4}\bigg)
\begin{pmatrix}
1\\
-1
\end{pmatrix}.
\end{dcases}
\end{equation}
We find that the optimal initial condition and the optimal readout are aligned respectively with the modes $(1,-1)$ and $(1,1)$. These modes correspond to the patterns of differential and equal firing of the excitatory and inhibitory units, respectively called the difference and sum modes in \citep{Murphy2009}. Thus, our theory recovers the results of \citep{Murphy2009}, showing that a difference in the firing of the E and I units drives strong changes in the pattern of common activation of E and I neurons.

\subsection{Appendix G}
The expression of the propagator and the mean activity are given by

\begin{equation}\label{eq-1}
\begin{dcases}
\big[e^{t(\Delta\uv-\I)}\big]_{ik}=e^{-t}(\delta_{ik}+\Delta t u_iv_k)\\
\langle r_i(t) \rangle=e^{-t}(v_i+\Delta t u_i).
\end{dcases}
\end{equation}
Using Eqs.~Eq.~\eqref{correlation-matrix} and Eq.~\eqref{eq-1} we can write the correlation matrix $\textbf{C}^g(t)$ as

\begin{equation}\label{correlation-matrix-t1}
\begin{split}
C_{ij}^g(t)=\frac{g^2}{N}e^{-2t}\sum_{k,l}&\int_0^t\deriv s_1  \big(\delta_{ik}+\Delta (t-s_1) u_iv_k\big)\big(v_l+\Delta s_1 u_l\big)
\int_0^t\deriv s_2  \big(\delta_{jk}+\Delta (t-s_2) u_jv_k\big)\big(v_l+\Delta s_2 u_l\big).
\end{split}
\end{equation}
By integrating over the variables $s_1$ and $s_2$ we find

\begin{equation}
\begin{split}
C_{ij}^g=\frac{g^2}{N}e^{-2t}&\sum_{k,l}\delta_{ik}\delta_{jk}v_l^2 t^2+\Delta\big(2\delta_{ik}\delta_{jk}v_lu_l+\delta_{ik}u_jv_kv_l^2+\delta_{jk}u_iv_kv_l^2\big)\frac{t^3}{2}\\
&+\Delta^2\big(\delta_{ik}\delta_{jk}u_l^2  \delta_{ik}u_ju_lv_kv_l+\delta_{jk}u_iu_lv_kv_l+u_iu_jv_k^2v_l^2\big)\frac{t^4}{4}+\Delta^2\big(\delta_{ik}u_ju_lv_kv_l+\delta_{jk}u_iu_lv_kv_l\big)\frac{t^4}{6}\\
&+\Delta^3\big(\delta_{ik}u_ju_l^2 v_k+\delta_{jk}u_iu_l^2 v_k+2u_iu_ju_lv_lv_k^2\big)\frac{t^5}{12}+\Delta^4u_iu_ju_l^2v_k^2\frac{t^6}{36}.
\end{split}
\end{equation}
By projecting $\textbf{C}^g$ on the direction $\U$, we find that only the order $1$, $\Delta^2$ and $\Delta^4$ contribute:

\begin{equation}
\U^T\textbf{C}^g\U=\sum_{i,j=1}^Nu_iC_{ij}(t)u_j=\frac{g^2}{N}e^{-2t}\Bigg( t^2+\Delta^2\frac{t^4}{2}+\Delta^4\frac{t^6}{36}\Bigg).
\end{equation}
At the time of the peak amplification, i.e. for $t=\tstar\simeq 1$, we recover Eq.~\eqref{correlation-matrix-t1}.

To compute the correlation matrix relative to the input noise, we use Eqs.~Eq.~\eqref{eq-1} and Eq.~\eqref{correlation-matrix}. As a result

\begin{equation}
C_{ij}^\sigma(t)=\sigma^2\bigg[\delta_{ij}\bigg(\frac{1}{2}-\frac{e^{-2t}}{2}\bigg)
+\Delta(u_iv_j+u_jv_i)\bigg(\frac{1}{4}-\frac{e^{-2t}}{4}-\frac{te^{-2t}}{2}\bigg)
+\Delta^2u_iu_j||\V||^2\bigg(\frac{1}{4}-\frac{e^{-2t}}{4}-\frac{te^{-2t}}{2}-\frac{t^2e^{-2t}}{2}\bigg)
\bigg].
\end{equation}
By projecting $\textbf{C}^\sigma$ evaluated at time $t=\tstar=1$ onto the readout $\U$, we find that only the order $1$ and $\Delta^2$ contribute, resulting in Eq.~\eqref{du_sigma}.

\subsection{Appendix H}

In this section we study the dynamics of the norm of the component of the activity $\textbf{r}(t)$ orthogonal to the plane defined by the two structure vectors $\U$ and $\V$. We focus on the case of uncorrelated structure vectors ($\rho\simeq 0$), so that the orthogonal component is given by $\textbf{r}^\perp(t)\simeq\textbf{r}(t)-\U\cdot\textbf{r}(t)-\V\cdot\textbf{r}(t)$. We assume that the condition for the strong amplification regime is satisfied (Eq.~\ref{strong_amplification_regime}) and we set the external input to the vector $\V$, which is close to the amplified initial condition in absence of noise in the connectivity ($g=0$). 

To study the temporal evolution of $||\textbf{r}^\perp||$, we project the dynamics onto a new orthonormal basis. We choose the first two basis vectors to be $\U$ and $\V$, while the choice of the remaining $N-2$ vectors is arbitrary, under the constraint that they form an orthonormal basis with $\U$ and $\V$. We call $\textbf{T}$ the orthogonal matrix which contains the new basis vectors as columns. The rate model in Eq.~\eqref{linear_rate_model} can be written in the new basis as

\begin{equation}\label{rank1_plus_noise_on_new_basis}
\dot{\tilde{\textbf{r}}}=-\tilde{\textbf{r}}+\tilde{\textbf{J}}\tilde{\textbf{r}}+\delta(t)\tilde{\textbf{r}}_0,
\end{equation}
where $\tilde{r}=(r_{\rm u},r_{\rm v},\textbf{r}^\perp)$, so that

\begin{equation}\label{orthogonal_norm}
||\textbf{r}^\perp(t)||=\sqrt{\tilde{r}_3^2(t)+...+\tilde{r}_N^2(t)}.
\end{equation} 
The connectivity matrix in the new basis is

\begin{equation}\label{new_connectivity_r1+noise}
\tilde{\textbf{J}}=\textbf{T}^T\J\textbf{T}\simeq
\begin{pmatrix}
\lambda & \Delta & \textbf{J}_{\rm u\perp}\\
0 & 0 & \textbf{J}_{\rm v\perp} \\
\textbf{J}_{\rm \perp u} & \textbf{J}_{\rm \perp v} &\textbf{J}_{\perp\perp}  \\
\end{pmatrix},
\end{equation}
where $\textbf{J}_{\rm \perp u}$ and $\textbf{J}_{\rm \perp v}$ are $(N-2)\times 1$ matrices, $\textbf{J}_{\rm u\perp}$ and $\textbf{J}_{\rm v\perp}$ are $1\times (N-2)$ matrices and $\textbf{J}_{\perp\perp}$ is a $(N-2)\times (N-2)$ matrix. Since $\textbf{T}$ and the connectivity noise $\bm{\chi}$ (see Eq.~\ref{rank-1_plus_noise}) are uncorrelated, the elements of these matrices have zero mean and variance equal to $g^2/N$. The elements $\tilde{\J}_{21}$ ans $\tilde{\J}_{22}$ are $O(1/\sqrt{N})$ and they have been set to zero in Eq.~\eqref{new_connectivity_r1+noise}. By differentiating both sides of Eq.~\eqref{orthogonal_norm} and using Eq.~\eqref{rank1_plus_noise_on_new_basis} and Eq.~\eqref{new_connectivity_r1+noise}, we can derive the equation for the dynamics of $||\textbf{r}^\perp(t)||$, which  reads:

\begin{equation}\label{R14}
\frac{\rm d\rPerp}{{\rm d}t}= \frac{\textbf{r}^{\perp T}(\textbf{J}_{\perp\perp,\,S}-1)\textbf{r}^{\perp}}{\rPerp^2}\rPerp
+\frac{\textbf{r}^\perp\cdot\textbf{J}_{\rm \perp v}}{\rPerp}\,\rv(t)
+\frac{\textbf{r}^\perp\cdot \textbf{J}_{\rm \perp u}}{\rPerp}\,\ru(t),
\end{equation}
where $\textbf{J}_{\perp\perp,\, S}$ denotes the symmetric part of $\textbf{J}_{\perp\perp}$. Eq.~\eqref{R14} alone is not enough to solve for the dynamics of $||\textbf{r}^\perp(t)||$, since it depends also on $\textbf{r}^\perp(t)$. However we note that, for $t\gg2/\Delta$ we have

\begin{equation}\label{R7}
\frac{\textbf{r}^\perp}{||\textbf{r}^\perp||}\simeq \textbf{J}_{\rm \perp u}.
\end{equation}
In fact, using Eq.~\eqref{rank1_plus_noise_on_new_basis} to compute the orthogonal activity for small times $\delta t$ we obtain

\begin{equation}
\textbf{r}^\perp(\delta t)
= \textbf{J}_{\rm \perp v}\delta t+\frac{1}{2}(\Delta\textbf{J}_{\rm \perp u}+
\textbf{J}_{\perp\perp}\textbf{J}_{\rm \perp v} )\delta t^2+O(\delta t^3).
\end{equation}
In the strong amplification regime (Eq.~\ref{strong_amplification_regime}), for times $\delta t\gg 2/\Delta$ we have $\Delta||\textbf{J}_{\rm \perp u}||\delta t^2\gg 2||\textbf{J}_{\rm \perp v}||\delta t+  ||\textbf{J}_{\perp\perp}\textbf{J}_{\rm \perp v}||\delta t ^2$, so that Eq.~\eqref{R7} holds up to corrections due to the input from the mode $\V$ and to the feedback from $\textbf{r}^\perp$ to itself. Numerical simulations confirm Eq.~\eqref{R7} and show that it holds also at larger times. The third term in Eq.~\eqref{R14} then becomes $g\ru(t)$.
Thus, neglecting  the second term on the right hand side of  Eq.~\eqref{R14}, which decays exponentially, and cosidering the mean activity along $\U$ given by Eq.~\eqref{solution_initial_condition_v}, we can write

\begin{equation}\label{eqforthenorm1}
\frac{\rm d\rPerp}{{\rm d}t}=-\gamma(t)\rPerp+g \Delta te^{-t},
\qquad
\gamma(t)=-\frac{\textbf{r}^{\perp T}(\textbf{J}_{\perp\perp,\,S}-1)\textbf{r}^{\perp}}{\rPerp^2}.
\end{equation}
Note that at time $t=0$ the elements of $\textbf{r}^\perp$ and $\textbf{J}_{\perp\perp,\,S}$ are uncorrelated, so that we have $\gamma(0)=1$. Instead, the asymptotic dynamics in the orthogonal subspace is governed by the coupling matrix $\textbf{J}_{\perp\perp}$ (see Eq.~\ref{new_connectivity_r1+noise}) so that the timescale of the decay of  $\rPerp$ is $1/(1-\lmax(\textbf{J}_{\perp\perp}))$, with $\lmax(\textbf{J}_{\perp\perp})=g-1$. Therefore the asymptotic value of $\gamma(t)$ is given by $\gamma(+\infty)=1-g$.
By solving Eq.~\eqref{eqforthenorm1} we obtain the expression  for the dynamics of $\rPerp$:

\begin{equation}
||\textbf{r}(t)||=g\Delta A\big(\gamma(g)\big),\quad A\big(\gamma(g)\big)=\int_0^t\deriv s\,\, se^{-\int_s^t\gamma(z)\deriv z\,-s}.
\end{equation}
Thus we find that, in presence of noise in the connectivity, the norm of the activity orthogonal to the $\UV$-plane scales linearly with $\Delta$.

\subsection{Appendix I}

The exponential of the sum of two matrices $\textbf{A}$ and $\textbf{B}$ can be factorized as

\begin{equation}
\exp(\textbf{A}+\textbf{B})=\exp(\textbf{A})\exp(\textbf{B})
\end{equation}
only if $\textbf{A}$ and $\textbf{B}$ commute, i.e. if the commutator $[\textbf{A},\textbf{B}]=\textbf{A}\textbf{B}-\textbf{B}\textbf{A}$ is equal to zero. In the following we compute the mean and the variance of the commutator 

\begin{equation}
C=[\Delta\U^{(1)}\V^{(1)T},\Delta\U^{(2)}\V^{(2)T}]
\end{equation}
and show that

\begin{equation}\label{AE3}
\langle C_{ij}\rangle=0,
\qquad
\langle C_{ij}^2\rangle\simeq \frac{2\Delta^4}{N^3}
\end{equation}
The mean of $C_{ij}$ is given by 

\begin{equation}
\begin{split}
\langle C_{ij}\rangle &=\sum_{k=1}^N\Big\langle \Uu_i\Vu_k\Ud_k\Vd_j- \Ud_i\Vd_k\Uu_k\Vu_j \Big\rangle.
\end{split}
\end{equation}
Since all the factors in the products on the right hand side are uncorrelated, we have $\langle C_{ij}\rangle=0$.
The variance of $C_{ij}$ is given by

\begin{equation}
\begin{split}
\langle C_{ij}^2\rangle=\sum_{k,l=1}^N&\Big\langle
\Uu_i\Vu_k\Ud_k\Vd_j\,\Uu_i\Vu_l\Ud_l\Vd_j +
\Ud_i\Vd_k\Uu_k\Vu_j\,\Ud_i\Vd_l\Uu_l\Vu_j \\
&-
\Uu_i\Vu_k\Ud_k\Vd_j\Ud_i\Vd_l\Uu_l\Vu_j
-\Ud_i\Vd_k\Uu_k\Vu_j\Uu_i\Vu_l\Ud_l\Vd_j
\Big\rangle.
\end{split}
\end{equation}
The first term on the right hand side is thus given by

\newcommand{\Uusq}{\U^{(1)2}}
\newcommand{\Vusq}{\V^{(1)2}}
\newcommand{\Udsq}{\U^{(2)2}}
\newcommand{\Vdsq}{\V^{(2)2}}

\begin{equation}\label{AE1}
\begin{split}
&\sum_{k,l=1}^N\Big\langle
\Uu_i\Vu_k\Ud_k\Vd_j\,\Uu_i\Vu_l\Ud_l\Vd_j\Big\rangle=
\sum_{k,l=1}^N \Big\langle \Uusq_i\Big\rangle 
\Big\langle \Vdsq_j\Big\rangle
\Big\langle \Vu_k\Vu_l\Big\rangle
\Big\langle \Ud_k\Ud_l\Big\rangle\\
&=\sum_{k,l=1}^N\frac{1}{N^4}\delta_{kl}=\frac{1}{N^3}.
\end{split}
\end{equation}
Computing the second term yields the same result. For the third term we obtain

\begin{equation}\label{AE2}
\begin{split}
&\sum_{k,l=1}^N\Big\langle
\Uu_i\Vu_k\Ud_k\Vd_j\Ud_i\Vd_l\Uu_l\Vu_j\Big\rangle=
\sum_{k,l=1}^N \Big\langle \Uu_i\Uu_l\Big\rangle 
\Big\langle \Vu_j\Vu_k\Big\rangle
\Big\langle \Ud_k\Ud_i\Big\rangle
\Big\langle \Vd_j\Vd_l\Big\rangle\\
&=\sum_{k,l=1}^N\frac{1}{N^4}\delta_{il}\delta_{jk}\delta_{ki}\delta_{jl}=\frac{1}{N^4}\delta_{ij}.
\end{split}
\end{equation}
Using Eq.~\eqref{AE1} and Eq.~\eqref{AE2}, we obtain Eq.~\eqref{AE3}. Thus, in the limit of large $N$ we can write

\begin{equation}
\exp\Big(t(\Delta\U^{(1)}\V^{(1)T}+\Delta\U^{2)}\V^{(2)T}-\I)\Big)=
e^{-t}\exp\Big(t(\Delta\U^{(1)}\V^{(1)T})\Big)\exp\Big(t(\Delta\U^{(2)}\V^{(2)T})\Big)
\end{equation}
and recover Eq.~\eqref{propagator_2_patterns}.

\clearpage

\section{Supplementary figures}

\renewcommand{\thefigure}{S\arabic{figure}}
\setcounter{figure}{0}

\begin{figure}[H]
\centering\includegraphics[scale=1]{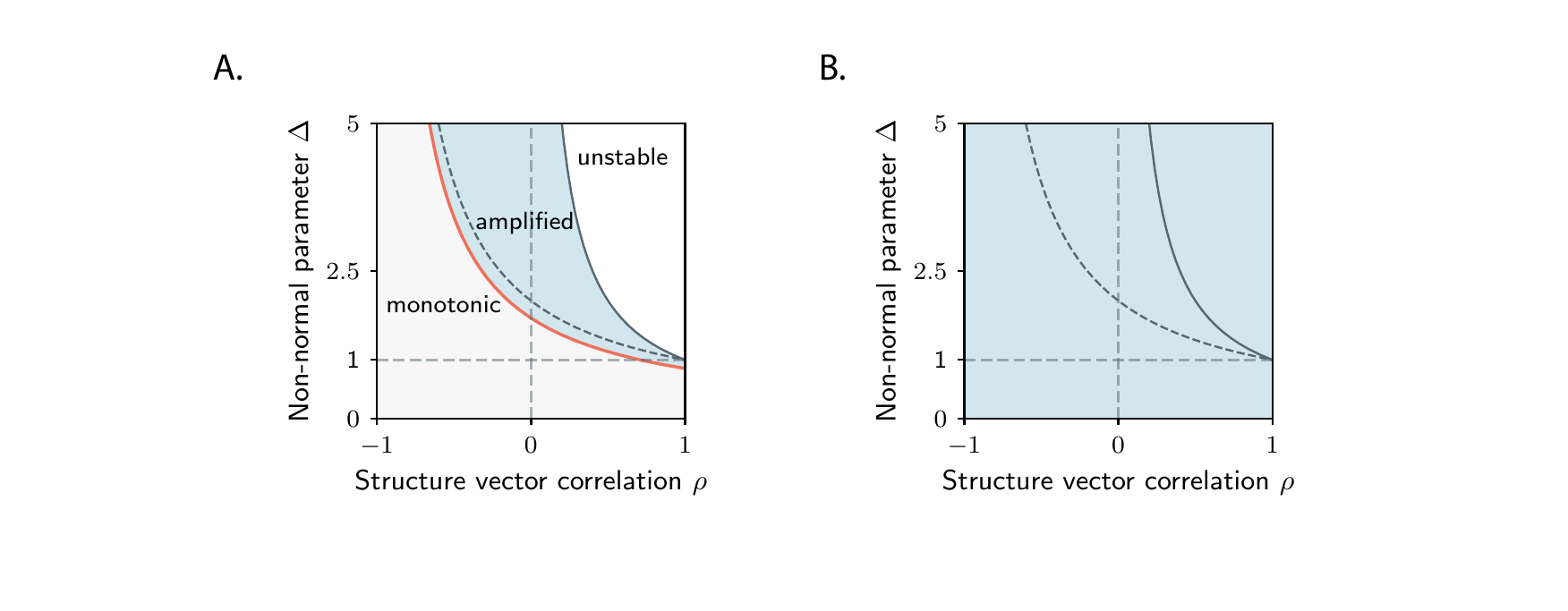} 
\caption{\label{fig:phase_diagram_noise}
\textsf{\bfseries Phase diagram for the unit-rank network with connectivity noise.}
\textsf{\bfseries A.} $g<1/\sqrt{2}$. The red line indicates the boundary between the monotonic and amplified parameter region for $g=0.5$. The grey dashed line corresponds to the case $g=0$. \textsf{\bfseries B.} $g>1/\sqrt{2}$. The dynamics are amplified regardless of the values of the parameters $\Delta$ and $\rho$.
}
\end{figure}

\begin{figure}[H]
\centering\includegraphics[scale=1]{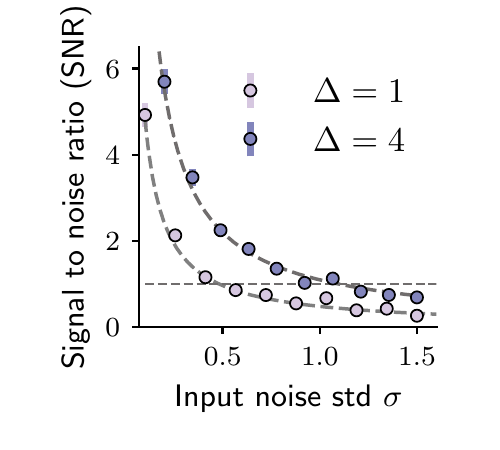} 
\caption{\label{fig:SNR}
Signal-to-noise ratio of the readout as a function of the standard deviation of the input noise $\sigma$ for two values of the non-normal parameter $\Delta$. Non-amplified dynamics ($\Delta=1$) are less robust to noise than amplified dynamics ($\Delta=4$). Dashed lines correspond to the theoretical values (Eq.~\ref{SNR}). In simulations, $N=1000$. Errorbars represent the standard deviation of the mean over $200$ realizations of the connectivity matrix.}
\end{figure}

\clearpage

\section{Acknowledgements}

We are grateful to Francesca Mastrogiuseppe and Manuel Beiran for discussions and feedback on the manuscript.

This work was funded by the Programme Emergences of City of Paris, Agence Nationale de la Rechere grant ANR-16-CE37-0016, and the program “Investissements d’Avenir” launched by the French Government and implemented by the ANR, with the references ANR-10- LABX-0087 IEC and ANR-11-IDEX-0001-02 PSL University. The funders had no role in study design, data collection and analysis, decision to publish, or preparation of the manuscript.

\addcontentsline{toc}{section}{References}
\bibliographystyle{abbrvnat}
\bibliography{bib2}

\end{document}